\documentclass[
reprint,
superscriptaddress,
nofootinbib,
 amsmath,amssymb,
 aps,
]{revtex4-2}

\usepackage{graphicx}
\usepackage[caption=false]{subfig}
\usepackage{dcolumn}
\usepackage{bm}
\usepackage{booktabs}
\usepackage{hyperref}
\hypersetup{colorlinks=true,linkcolor=blue,citecolor=blue,urlcolor=blue}
\usepackage{orcidlink}
\defcitealias{Arnaud_2010}{A10}

\usepackage{xcolor}

\begin{document}


\title{Impact of non-Gaussian likelihood on cosmological constraints from the thermal Sunyaev--Zel'dovich power spectrum: a simulation-based inference analysis}

\author{Licong Xu\,\orcidlink{0009-0005-7088-4379}}
\email{lx256@cam.ac.uk}   
\affiliation{Institute of Astronomy, University of Cambridge, Madingley Road, Cambridge CB3 0HA, UK}
\affiliation{Kavli Institute for Cosmology, University of Cambridge, Madingley Road, Cambridge CB3 0HA, UK}

\author{Íñigo Zubeldia\,\orcidlink{0000-0002-1879-4289}}
\affiliation{Institute of Astronomy, University of Cambridge, Madingley Road, Cambridge CB3 0HA, UK}
\affiliation{Kavli Institute for Cosmology, University of Cambridge, Madingley Road, Cambridge CB3 0HA, UK}
\affiliation{DAMTP, Centre for Mathematical Sciences, Wilberforce Road, Cambridge CB3 0WA, UK}

\author{James Alvey\,\orcidlink{0000-0003-2020-0803}}
\affiliation{Institute of Astronomy, University of Cambridge, Madingley Road, Cambridge CB3 0HA, UK}
\affiliation{Kavli Institute for Cosmology, University of Cambridge, Madingley Road, Cambridge CB3 0HA, UK}

\author{Boris Bolliet\,\orcidlink{0000-0003-4922-7401}}
\affiliation{Cavendish Astrophysics, University of Cambridge, Madingley Road, Cambridge CB3 0HA, UK}
\affiliation{Kavli Institute for Cosmology, University of Cambridge, Madingley Road, Cambridge CB3 0HA, UK}

\author{Anthony Challinor\,\orcidlink{0000-0003-3479-7823}}
\affiliation{Institute of Astronomy, University of Cambridge, Madingley Road, Cambridge CB3 0HA, UK}
\affiliation{Kavli Institute for Cosmology, University of Cambridge, Madingley Road, Cambridge CB3 0HA, UK}
\affiliation{DAMTP, Centre for Mathematical Sciences, Wilberforce Road, Cambridge CB3 0WA, UK}

\date{\today}

\begin{abstract}
    The thermal Sunyaev--Zel'dovich (tSZ) power spectrum is a sensitive probe of cosmology and cluster astrophysics, but its statistics are non-Gaussian because the signal receives a significant contribution from rare, massive, low-redshift galaxy clusters.
    As a result, a Gaussian likelihood fails to describe the statistics of its power spectrum on large scales. We use simulation-based inference (SBI) to test the accuracy of the standard Gaussian power-spectrum likelihood for a \textit{Planck}-like tSZ analysis.
    Using halo-based simulations of full-sky Compton-$y$ maps, we train neural posterior and likelihood estimators and compare the resulting constraints with those from a Gaussian likelihood assumption. 
    Using only multipoles $\ell < 1000$, we find that the Gaussian likelihood assumption gives unbiased cosmological constraints, while the SBI-based inference shows a mild broadening of the posterior distributions for the amplitudes of residual foregrounds.
    This suggests that the Gaussian likelihood assumption is sufficiently accurate for cosmological inference for a \textit{Planck}-like tSZ analysis, while SBI provides a useful validation tool to model non-Gaussian likelihoods beyond analytic approximations.
    \end{abstract}

\maketitle


\section{Introduction}
Galaxy clusters are the largest gravitationally bound structures in the Universe.
They form in the most massive dark-matter halos, with typical masses of a few $10^{14}\,M_\odot$. The abundance of galaxy clusters is highly sensitive to the cosmological model and the growth of cosmic structure across time. This has motivated numerous cosmological analyses using cluster number counts. In the past, the properties of galaxy clusters have been determined through different observations at different wavelengths, ranging from X-ray (see, e.g., \citep{Clerc_2012} for a review), and optical \citep{DES:2017myr, KiDS:2020suj,LSSTDarkEnergyScience:2018jkl} to Sunyaev--Zel'dovich (SZ) surveys \citep{ade2013planck,Salvati:2017rsn,George:2014oba, Zubeldia2024b}. 


The thermal Sunyaev--Zel'dovich (tSZ) effect~\citep{1972CoASP...4..173S} is one of the most important probes to study galaxy clusters. It produces a spectral distortion of the cosmic microwave background (CMB) due to the inverse Compton scattering of the CMB photons by the hot electrons in the intracluster medium (ICM).
Detecting such distortions allows one to construct tSZ maps tracing the large-scale distribution of hot gas. Cosmological and astrophysical information can then be extracted from these maps using a range of summary statistics (see, e.g., \citep{Sabyr_2025} for a review), and using cluster number counts derived from the tSZ signal (see, e.g., \citep{Ade2016, Zubeldia2019, Bocquet2024b, Zubeldia:2025qlt}). 

Among all the summary statistics that can be extracted from tSZ maps, the power spectrum is one of the simplest and most direct probes of cosmological information. On large scales, the tSZ power spectrum depends strongly on the matter clustering parameter $\sigma_8$, and matter density fraction $\Omega_m$ \citep{Komatsu_1999,Komatsu:2002wc}, and on the non-linear gas physics within the clusters on small scales \citep{McCarthy_2014}.
In particular, the tSZ power spectrum contains information about all the clusters that are not detected individually in cluster number-count analyses, and thus offers a complementary probe to the cluster number-count method \citep{Rotti:2020rdl}.


Previous analyses of the tSZ power spectrum have typically assumed a Gaussian likelihood for the power spectrum. For example, the \citet{Planck:2015vgm} analysis constrained both the cosmological parameters and residual foreground amplitudes, using a covariance based only on the Gaussian contribution.
\citet{Bolliet:2017lha} revisited this approach by incorporating the non-Gaussian trispectrum contribution to the covariance, and by varying all relevant cosmological and foreground parameters simultaneously. More recent analyses include a \textit{Planck} tSZ power spectrum analysis based on the PR4 data \citep{Tanimura_2021}, and a measurement of the full shape of the tSZ power spectrum from the South Pole Telescope \citep{SPTpol:2026xmb}. However, the contribution to the tSZ power spectrum is dominated by massive clusters, whose rarity introduces non-Gaussianity. This effect is more significant on large angular scales, where the signal is dominated by a small number of the largest clusters, and is less prominent on smaller scales as the contribution becomes distributed over a large number of less-massive clusters. Several studies (e.g.,~\citep{Zhang:2007psa,2009MNRAS.397.2189P}) have shown that the sampling distribution of the tSZ power spectrum is non-Gaussian and positively skewed at low multipoles even for large sky fractions, approaching Gaussianity only at higher multipoles and for large sky fractions. Importantly, this type of non-Gaussianity is beyond that due to the finite-mode statistics of the power spectrum estimator at low multipoles.
Nevertheless, a likelihood model that accurately captures the non-Gaussian statistics of the tSZ power spectrum remains unavailable.

In this work, we propose an alternative method to obtain cosmological constraints from the tSZ power spectrum using a simulation-based inference (SBI) approach (see, e.g., \citep{Cranmer_2020} for a comprehensive review).
SBI enables parameter inference directly from forward-modeled simulations, without requiring an explicit analytic likelihood. Working in a \textit{Planck}-like setting, we generate mock galaxy cluster catalogs and produce their corresponding sky maps for a given set of cosmological parameters, including residual foreground contamination. We then train deep neural networks to learn either the posterior distribution directly via Neural Posterior Estimation (NPE;~\citep{papamakarios2018fastepsilonfreeinferencesimulation}) or to learn the likelihood function through Neural Likelihood Estimation (NLE;~\citep{papamakarios2019sequentialneurallikelihoodfast}).
We validate our SBI-based approach using mock sky maps generated across a range of input parameters with the same pipeline as that used for training, and compare the consistency between the learned likelihood and the true sampling distribution. 

This paper is organized as follows. In Secs~\ref{sec:theorytsz} and~\ref{stattsz}, we review the theoretical calculation of the tSZ power spectrum within the halo-model formalism and its statistical properties. Next, in Sec.~\ref{sec:lklformalism}, we outline the Gaussian-likelihood formalism of the tSZ power spectrum for constraining parameters. Sec.~\ref{sec:halobasedsims} describes our forward-modeling approaches for the tSZ power spectrum, including both a Gaussian-likelihood-based simulator and the halo-based simulations that generate full-sky maps. In Sec.~\ref{sec:sbi}, we present our SBI-based approach and the details of the neural density estimation implementations. The main parameter results are given in Sec.~\ref{sec:results}, and we validate our SBI-based approach in Sec.~\ref{subsec:validation_sbi}. We conclude with a discussion and outlook in Sec.~\ref{sec:discussion}. 

\begin{figure}
    \centering
    \includegraphics[width=0.99\linewidth]{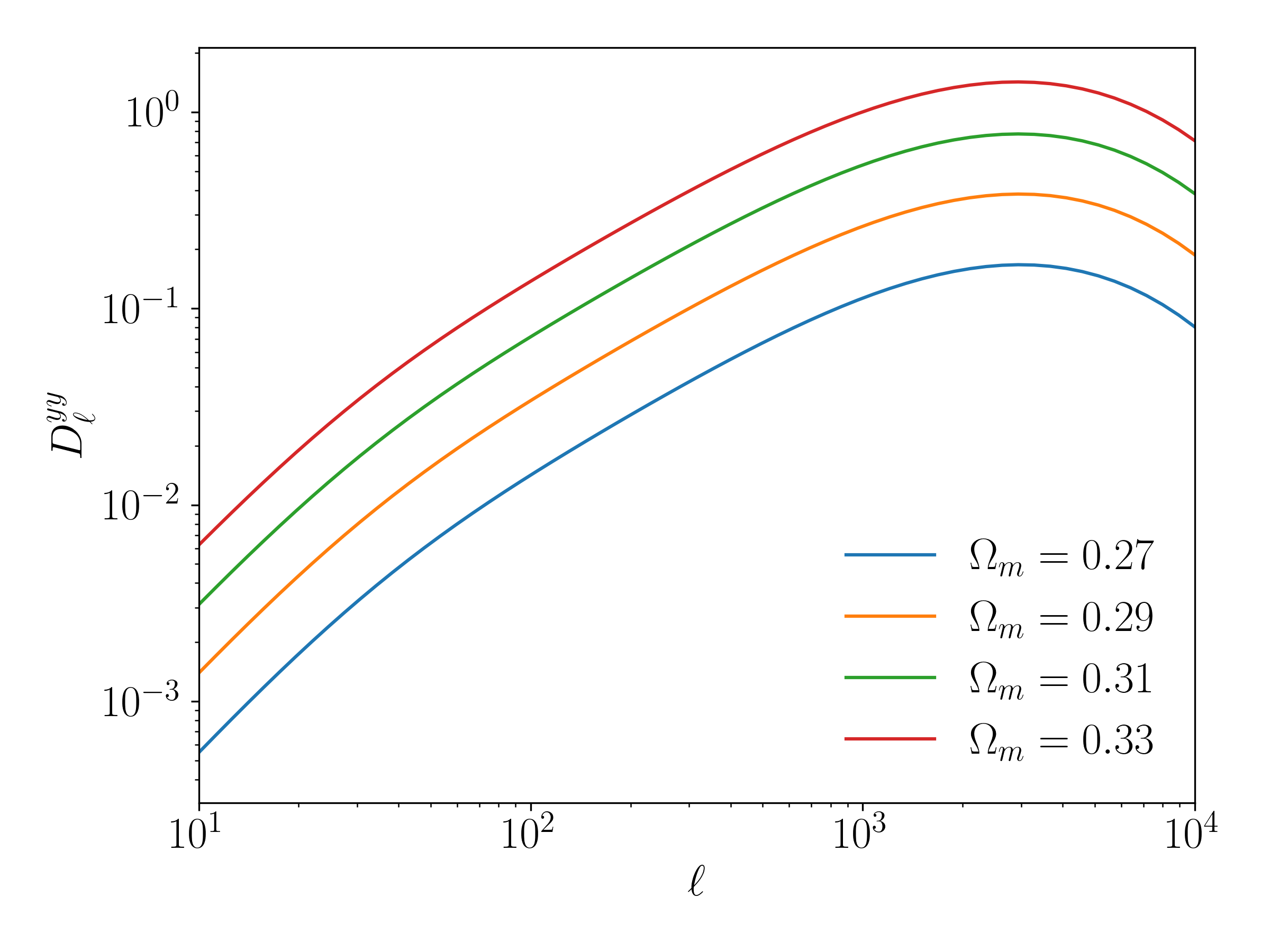}\\[4pt]
    \includegraphics[width=0.99\linewidth]{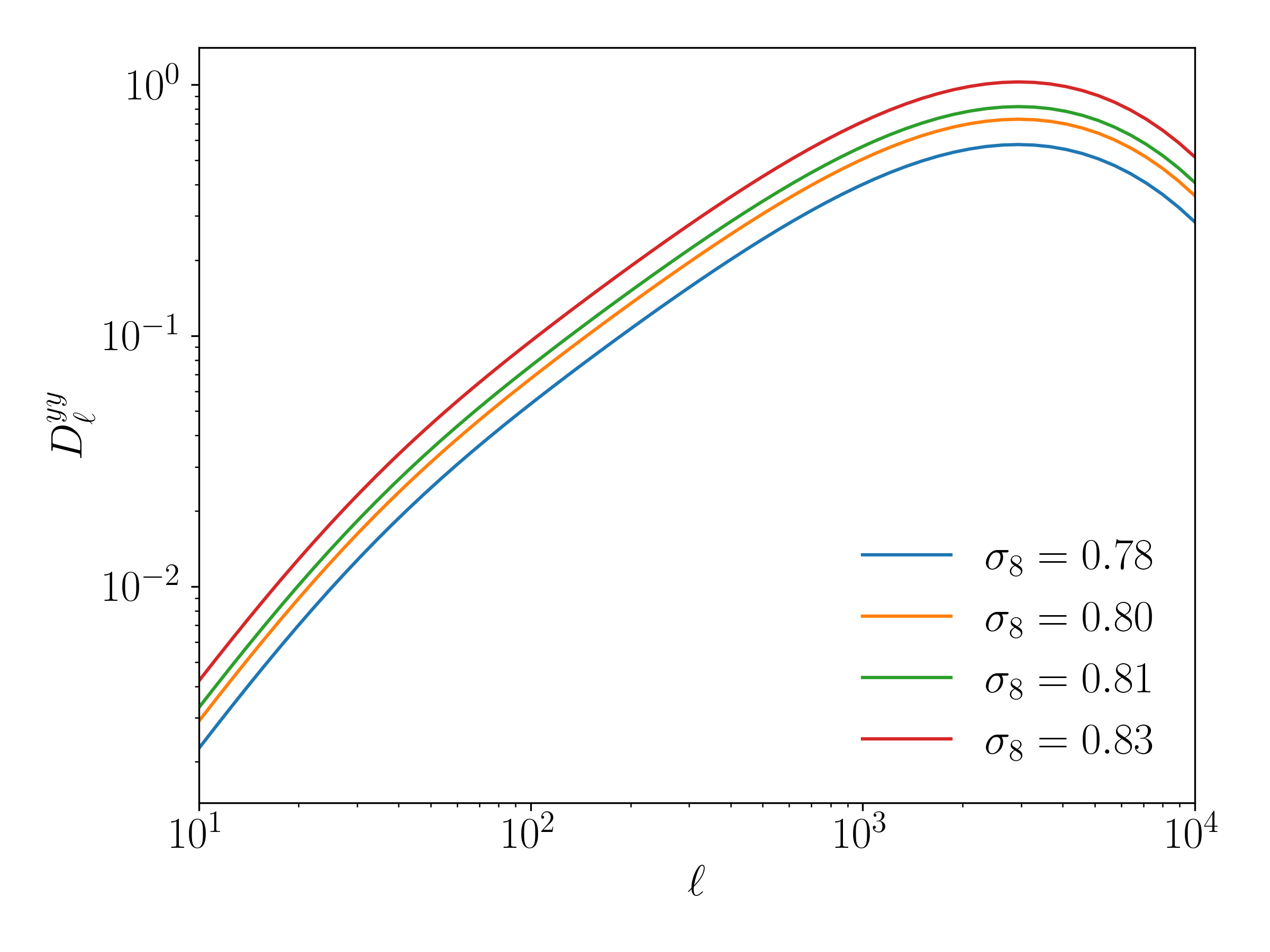}
    \caption{
        Dependence of the tSZ power spectrum, $D_\ell^{yy} = 10^{12} \times \ell(\ell+1)C_\ell^{\rm tSZ}/{(2\pi)}$, on $\Omega_m$ (upper) and $\sigma_8$ (lower) with fixed $\omega_\text{b}$, $h$ and $n_\text{s}$.
        }
        

    \label{fig:tszps_omega_sigma}
\end{figure}

\section{Theory of the tSZ power spectrum}
\label{sec:theorytsz}

\subsection{Halo-model formalism of the tSZ power spectrum}
The tSZ effect introduces a frequency-dependent shift of the observed CMB temperature. In the non-relativistic limit, it can be written as~\citep{1972CoASP...4..173S}
\begin{equation}
    \frac{\Delta T(\nu)}{T_{\rm CMB}} = f(\nu)y(\theta),
    \label{tSZtemp}
\end{equation}
where $\Delta T(\nu)$ is the frequency-dependent temperature deviation, $T_{\rm CMB} = 2.7255\,\text{K}$ is the temperature of the undistorted CMB~\citep{Fixsen_2009}, $f(\nu) = x {\rm coth}(x/2) -4$ is the tSZ spectral function with $x =h_{\rm P} \nu/(k_{\rm B} T_{\rm CMB})$ and $h_{\rm P}$, $k_{\rm B}$ are the Planck and Boltzmann constants, respectively. 
Also, $y(\theta)$ is known as the Compton-$y$ parameter, which is the line-of-sight integral of electron pressure as a function of angular position $\theta$:
\begin{eqnarray}
        y(\theta) = \frac{\sigma_{\text{T}}}{m_{e} c^{2}}\int P_{e}\left(\sqrt{l^{2} + d_{A}^{2}\theta^{2}}
        \right) \, dl ,
\label{comptonyint}
\end{eqnarray}
where $\sigma_{\text{T}}$ is the Thomson cross-section, $m_{e}$ is the electron mass, $c$ is the speed of light, $P_{e}$ is the electron pressure, 
$l$ is the line-of-sight distance, and $d_{A}$ is the angular diameter distance.

We adopt a halo-model formalism \citep{Cooray:2002dia} to model the tSZ power spectrum. In this formalism, the tSZ power spectrum can be written as the sum of 1-halo and 2-halo terms. The 1-halo term
is the sum of the squared two-dimensional Fourier transform of the pressure profile from individual clusters.
In the continuous limit, the 1-halo term can be written as
\begin{align}
C_{\ell}^{\rm tSZ, \rm 1h}&= \int_{0}^{z_{\rm max}}dz \, \frac{dV}{dz\,d\Omega} \nonumber \\
        & \hspace{1cm}\times \int_{M_{\rm min}}^{M_{\rm max}}dM\, \frac{dn(M,z)}{dM}\, \left| \tilde y_{\ell}
        (M,z) \right|^{2}, 
\label{eq:tsz_power_spectrum}
\end{align}
where $dV/(dz d \Omega)$ is the derivative of comoving volume $V$ with respect to redshift $z$ and solid angle $\Omega$ 
, and $dn/dM$ is the halo mass function (HMF), for which we adopt the form of  Tinker \textit{et al.}~(2008) \citep{Tinker:2008ff}. 
The quantity $\tilde y_\ell$ is the two-dimensional Fourier transform of the (spherically symmetric) Compton-$y$ profile, $y(\theta)$, for a cluster of mass $M$ and redshift $z$ given by
\begin{equation}
        \tilde y_{\ell}(M,z) = \frac{4\pi r_{s}}{\ell_{s}^{2}}\frac{\sigma_{T}}{m_{e} c^{2}}
        \int_{0}^{\infty} dx\, x^{2}\, \frac{\sin(\ell x/\ell_{s})}{\ell x/\ell_{s}}
        \, P_{e}(x;M,z),
\end{equation}
where $x=r/r_s$, with $r$ the radial distance to the center of the halo and $r_s$ the characteristic scale radius, and $\ell_s = d_A/r_s$. In this paper, we set $r_s=r_{500c}/c_{500}$, where $r_{500c}$ is the radius within which the mean enclosed density is 500 times the critical density of the universe at the cluster redshift,
and $c_{500}$ is the corresponding concentration parameter. The 2-halo term describes the correlation between spatial positions of the clusters, and is relevant on large angular scales. The 2-halo term is not significant compared to the 1-halo term when all the massive and local clusters are included,
but can have a non-negligible contribution when those clusters are masked out \citep{Komatsu_1999}. We neglect the 2-halo term in this paper since we include the rare massive and local clusters in our analysis. 

The electron pressure profile is assumed to follow the generalized Navarro--Frenk--White (GNFW) profile given by
\begin{equation}
        P_{e}(x) = C \times P_{0} (c_{500}x)^{-\gamma}[1 + (c_{500}x)^{\alpha}]^{(\gamma-\beta)/\alpha},
\end{equation}
where, recall, $x=r/r_s$.
In this paper, we adopt the pressure profile of  \citeauthor{Arnaud_2010}~(\citeyear{Arnaud_2010}) \citep{Arnaud_2010}, hereafter~\citetalias{Arnaud_2010}, where the parameters $\{\gamma, \alpha, \beta, P_0, c_{500}\}$ are set to be constants, and the coefficient $C$ is given by
\begin{align}
        C &= 1.65 \left(\frac{h}{0.7}\right)^{2}\ \left(\frac{H(z)}{H_{0}}\right)^{8/3}
        \nonumber \\
        & \quad \times \left(\frac{\tilde{M}_{500c}}{0.7 \times 3 \times 10^{14}\,M_{\odot}}\right)^{2/3+0.12}
        \;\mathrm{eV\,cm^{-3}}.
\end{align}
%
Here, $H(z)$ is the Hubble parameter, with present-day value $H_0 = 100 h\,\text{km}\,\text{s}^{-1}\text{Mpc}^{-1}$. The mass $\tilde M_{500c}$ is related to the true mass via the relation $\tilde M_{500c} = M_{500c}/B$, where $M_{500c}$ is the mass enclosed within a radius inside which the mean density is 500 times the critical density of the universe at redshift $z$, and the bias factor $B$ accounts for the difference between X-ray mass proxies and the true halo mass that affects the scaling relations. In the literature, a different variable $1-b$ is used more often and it is related to $B$ via $B=1/(1-b)$.
This can be caused by non-hydrostatic pressure support, such as turbulence, magnetic fields, or cosmic rays \citep{Rasia:2014sga}, instrumental calibration, or selection effects \citep{2012MNRAS.426.2046A}. Since $r_{500c}$ is defined via the mass estimates,  a biased mass estimate $\tilde M_{500c} = M_{500c}/B$ in the pressure profile scaling relation also changes the corresponding scale radius, giving $r_{500c}\propto(M_{500c}/B)^{1/3}$.


Throughout the paper, we assume a flat $\Lambda$CDM cosmology with an effective number of neutrino species $N_{\rm eff}=3.046$, and the sum of neutrino masses to be $\sum m_\nu = 0.06$ eV. Unless otherwise stated, all benchmark cluster catalogs and Compton-$y$ maps are produced at fiducial cosmological parameters defined by $\omega_{\rm b} = 0.02242$, $\omega_{\rm cdm} = 0.1193$, $H_{0} = 67.66~{\rm km\,s^{-1}\,Mpc^{-1}}$, $\ln(10^{10}A_{\rm s}) = 2.9718$, $n_{\rm s} = 0.9665$, and a fiducial hydrostatic bias parameter $B = 1.41$, which we take to be independent of mass and redshift. 

\subsection{Parameter dependence of the tSZ power spectrum}
On large scales, the shape of the tSZ power spectrum is sensitive to the matter density, $\Omega_m$, and amplitude of matter fluctuations, $\sigma_8$. Figure~\ref{fig:tszps_omega_sigma} illustrates how variations in $\Omega_m$ and $\sigma_8$ affect the shape of the tSZ power spectrum. For the multipole range relevant to the Planck survey ($\ell < 10^3$), \citet{Komatsu:2002wc} find that the tSZ power spectrum scales approximately as 
\begin{equation}
            C_{\ell}^{\rm tSZ} \propto \sigma_{8}^{8.1} \, \Omega_{m}^{3.2} \, B^{-3.2} \, h^{-1.7}.
\end{equation}
Hence, the tSZ power spectrum is sensitive to the parameter combination $F$ defined as $F=\sigma_8(\Omega_m/B)^{0.40}h^{-0.21}$. 

        


\subsection{Mass and redshift dependence of the tSZ power spectrum}
\label{tszmzdep}
We now investigate the halo masses and redshifts to which the tSZ power spectrum is most sensitive at $\ell \leq 10^3$. The differential contributions of mass and redshift to the 1-halo tSZ power spectrum follow from Eq.~\eqref{eq:tsz_power_spectrum}.
We have
\citep{Komatsu:2002wc}: 
\begin{align}
\frac{d \ln C_\ell^{\rm tSZ}}{d \ln M} &\equiv 
\frac{
    M \int dz \frac{dV}{dzd\Omega} \frac{dn(M, z)}{dM} \left| \tilde{y}_\ell(M, z) \right|^2
}{
    \int dz \frac{dV}{dzd\Omega} \int dM \frac{dn(M, z)}{dM} \left| \tilde{y}_\ell(M, z) \right|^2
}, \\
\frac{d \ln C_\ell^{\rm tSZ}}{d \ln z} &\equiv 
\frac{
    z \frac{dV}{dzd\Omega} \int dM \frac{dn(M, z)}{dM} \left| \tilde{y}_\ell(M, z) \right|^2
}{
    \int dz \frac{dV}{dzd\Omega} \int dM \frac{dn(M, z)}{dM} \left| \tilde{y}_\ell(M, z) \right|^2
}.
\end{align}



\begin{figure}
    \centering
    \includegraphics[width=0.99\linewidth]{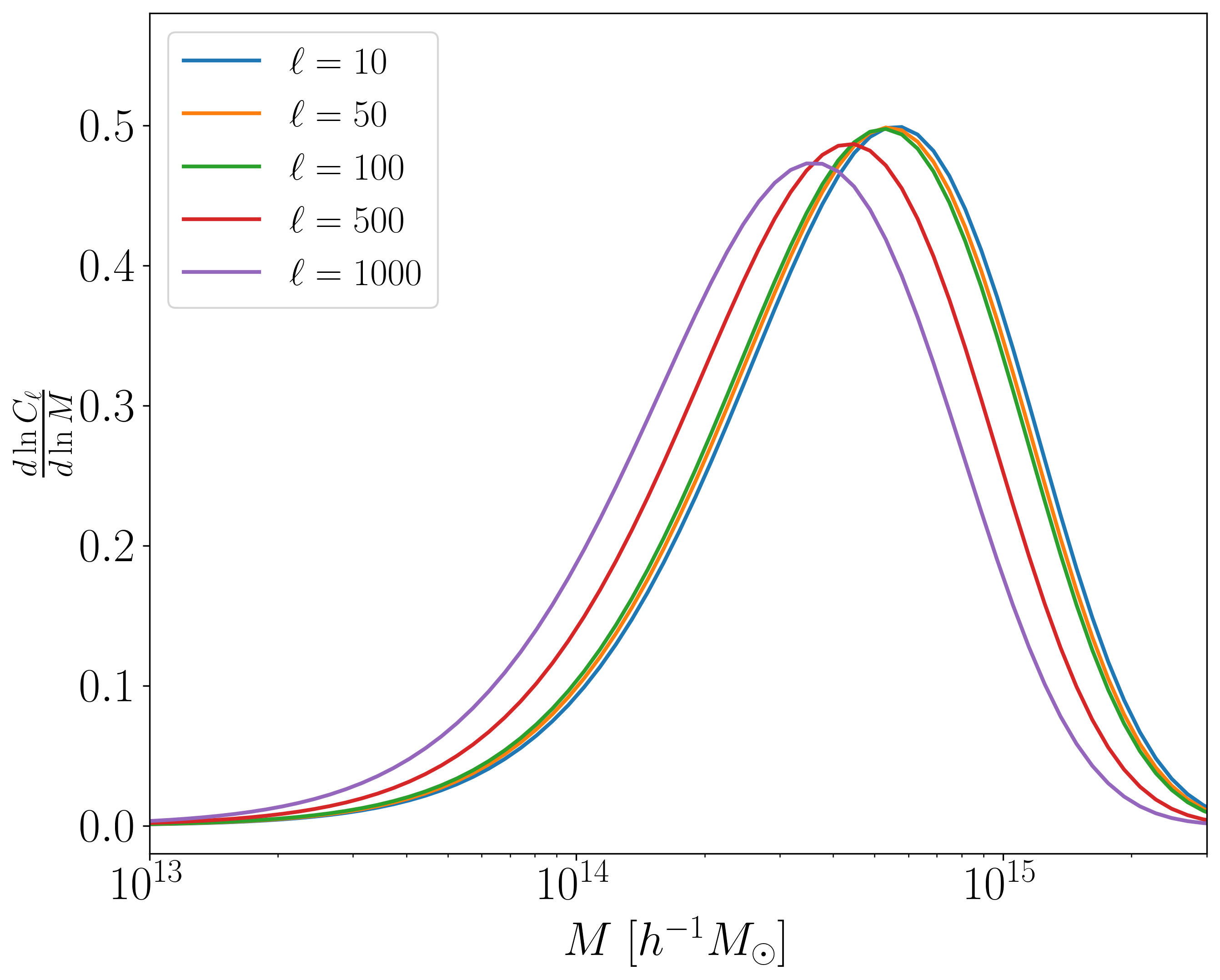}
    \includegraphics[width=0.99\linewidth]{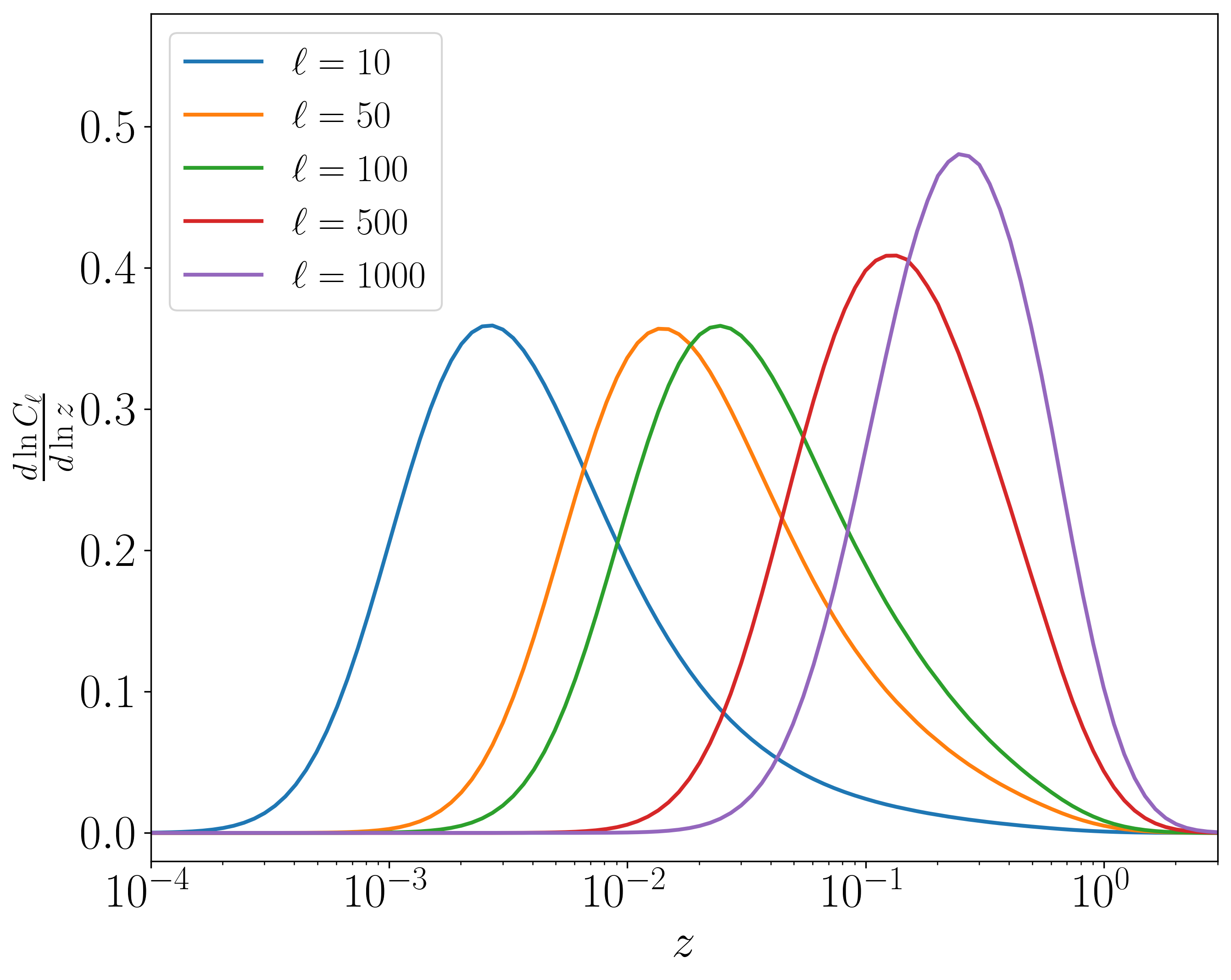}
    \caption{Upper panel: Differential mass contribution to $C_\ell^{\rm tSZ}$, $d \ln C_\ell^{\rm tSZ}/d\ln M$, for $\ell = 10$, $50$, $100$, $500$, and $1000$. Lower panel: Differential redshift contribution to $C_\ell^{\rm tSZ}$, $d \ln C_\ell^{\rm tSZ}/d\ln z$, for the same multipoles.}
    \label{fig:dcldmz}
\end{figure}

Figure~\ref{fig:dcldmz} shows the differential mass and redshift distribution of $C_\ell^{\rm tSZ}$ for multipoles $\ell=10$, 50, 100, 500, and 1000.  We find that the contribution to the tSZ power spectrum gradually shifts toward lower-mass halos as $\ell$ increases, since higher multipoles probe smaller angular scales corresponding to more numerous small halos. At $\ell=1000$, the mass distribution of $C_\ell^{\rm tSZ}$ peaks at $M \sim 4\times10^{14}\, h^{-1}M_\odot$. It is the massive halos between $10^{14}\,h^{-1}M_\odot$ and $10^{15}\, h^{-1}M_\odot$ that dominate $C_\ell^{\rm tSZ}$ over this multipole range of interest. At large angular scales $(\ell \lesssim 100)$, the distribution converges, indicating that these modes are already dominated by the most massive clusters. In this regime, further increasing the angular scale produces negligible changes in the range of halo masses contributing to the signal. In addition, the tSZ power spectrum is dominated by low-redshift halos for all these multipoles. The halos at $z\sim10^{-2}$ dominate the contribution to $C_\ell^{\rm tSZ}$ at $\ell \sim 100$, and halos at $z \sim 10^{-1}$ contribute to $C_\ell^{\rm tSZ}$ at $\ell \sim 1000$. Halos with $z>2$ have a negligible contribution to $C_\ell^{\rm tSZ}$ for $\ell < 1000$,
as their tSZ signal is strongly suppressed by the decrease in halo abundance and angular size with redshift.


\section{Statistics of the tSZ power spectrum}
\label{stattsz}

In this section, we discuss the sampling distribution of the tSZ power spectrum, both analytically and through simulations. We find that the distributions are significantly skewed due to rare, massive and local clusters, calling into question the usual Gaussian likelihood approximation.

We start by revisiting the derivation in Ref.~\cite{Zhang:2007psa} of the $n$-point joint probability distribution function (PDF) of the 1-halo term of the tSZ power spectrum. We extend it by including the scatter of the tSZ power spectrum given a fixed set of clusters on the sky due to their random locations.


Suppose there are $N$ clusters contributing to the tSZ power spectrum, and we bin these into $\alpha$ bins (e.g., mass, redshift and solid angle). Let $N_i$ be the number of clusters in the $i$th bin.
The 1-point distribution of the tSZ power spectrum $C$ at specific multipole $\ell$, $P(C)$, can be written as
\begin{equation}
P(C)=\sum_{N_1=0}^{\infty}\ldots\sum_{N_\alpha=0}^{\infty}P(N_1,\ldots,N_\alpha)P(C|N_1, \ldots, N_\alpha).
\end{equation}
Assuming that clusters are drawn from the HMF according to a Poisson process,
the number of clusters in each bin is independent of one another. The joint probability $P(N_1,\ldots,N_\alpha)$ is then given by 
\begin{equation}
    P(N_1,\ldots,N_\alpha) = \prod_{i=1}^\alpha \frac{\lambda_i^{N_i}}{N_i!} \exp(-\lambda_i),
\end{equation}
where $\lambda_i$ is the mean number of clusters in each bin $i$. We also assume that for a given set of clusters $\{N_i\}$, the tSZ power spectrum has some scatter around the mean, which is assumed to be Gaussian with variance $\sigma_G^2$. In principle, $\sigma_G^2$ is a function of the number of clusters $N_i$ as shown in Appendix \ref{app:conditional_scatter}.
We further assume that for typical $N_i$, any dependence of $\sigma_G^2$ on them may be ignored. Hence, $P(C|N)$ can be written as
\begin{equation}
    P(C|N) = \frac{1}{\sqrt{2\pi \sigma_G^2}} \exp\!\left[-\frac{(C - \bar{C})^2}{2\sigma_G^2}\right],
\end{equation}
where the mean is given by the 1-halo contribution from individual clusters, $\bar{C}=\frac{1}{\Omega_{\rm sky}}\sum_{i=1}^\alpha N_i S_i$, with $S_i = |\tilde{y}_\ell(M_i,z_i)|^2$ being the square of the 2D Fourier transform of the Compton-$y$ profile of the clusters in the $i$th bin.
Also, $\Omega_{\rm sky}=4\pi f_{\rm sky}$ is the sky area in steradians and $f_{\rm sky}$ is the sky fraction. 

To evaluate the probability density function (PDF) of $C$, $P(C)$, it is convenient to consider the characteristic function of $P(C)$,
\begin{align}
    \tilde{P}(\omega) 
    &= \int P(C)\, \exp\!\left(-i\omega C\right)\, dC \nonumber\\
    &= \exp\!\left\{ \sum_{i=1}^\alpha \lambda_i 
        \left[ \exp\!\left(-i\omega S_i/\Omega_{\rm sky}\right) - 1 \right] \right\} \nonumber \\
       & \hspace{1cm} \times \exp\!\left(-\frac{\sigma_G^2 \omega^2}{2}\right).
\end{align}
In the continuous limit where the bin size goes to zero, the characteristic function becomes
\begin{equation}
    \ln   \tilde{P}(\omega) = \int d\bar{N}  
        \left[ \exp\!\left(-i\omega S/\Omega_{\rm sky}\right) - 1 \right]  -\frac{\sigma_G^2\omega^2}{2}, 
\label{1ptpdfgeneral}
\end{equation}
where the first term characterizes the Poisson process of cluster number counts modulated by the Compton-$y$ profile of the clusters, and the second term accounts for the Gaussian scatter of the power spectrum.

Suppose we bin the clusters in bins of mass, redshift, and solid angle, $(M,z,\Omega)$. Then, in our notation, Eq.~\ref{1ptpdfgeneral} can be written as
\begin{align}
\ln \tilde{P}(\omega) &=
\Omega_{\rm sky} \int dz\, \frac{dV}{dz\,d\Omega}
\int dM \, \frac{dn}{dM}
\nonumber \\
& \hspace{1cm}\times \left[ \exp\!\left(-i\omega 
\frac{|\tilde{y}_\ell(M,z)|^{2}}{\Omega_{\rm sky}}\right) - 1 \right]
\nonumber \\
& \hspace{1cm} - \frac{\sigma_G^2 \omega^2}{2}.
\label{1ptpdfourcase}
\end{align}

The formalism above can be easily generalized to the $n$-point joint PDF of the tSZ power spectrum, $P(C_1, ..., C_n)$, by considering the joint characteristic function, 
%
\begin{align}
    \ln \tilde{P}(\omega_1, \ldots, \omega_n) 
    &= \Omega_{\text{sky}} \int dz \, \frac{dV}{dz\,d\Omega} 
    \int dM \, \frac{dn}{dM} \nonumber \\
    &\quad \times
    \Bigg\{ 
    \exp\!\left[-i \sum_{i=1}^{n} 
    \omega_i \frac{|\tilde{y}_{\ell_i}(M,z)|^2}{\Omega_{\rm sky}}\right]
    - 1 \Bigg\} \nonumber \\
    &\quad - \frac{1}{2} 
    \sum_{i=1}^{n} \omega_i^2 \, \sigma_{i, \rm G}^{2}.
\label{nptjointpdf}
\end{align}
\begin{widetext}
Here, $\tilde{y}_{\ell_i}(M,z)$ is the two-dimensional Fourier transform of the cluster Compton-$y$ profile at multipole $\ell_i$ in Eq.~\ref{nptjointpdf}, and $\sigma_{i,{\rm G}}^2$ denotes the Gaussian contribution to the variance in the corresponding multipole bin.
The connected moments of the $C_i$ follow from $\ln \tilde{P}(\omega_1,\ldots,\omega_n)$ by the cumulant expansion theorem. Expanding the right-hand side of Eq.~\ref{nptjointpdf} in powers of $\omega$, gives
\begin{equation}
\label{eq:expansion}
\exp\!\left[-i \sum_{i=1}^{n} \omega_i \frac{|\tilde{y}_{\ell_i}(M,z)|^2}{\Omega_{\rm sky}}\right] - 1 = -i \sum_{i=1}^{n} \omega_i \frac{|\tilde{y}_{\ell_i}(M,z)|^2}{\Omega_{\rm sky}} - \frac{1}{2} \sum_{i=1}^{n} \sum_{j=1}^{n} \omega_i \omega_j \frac{|\tilde{y}_{\ell_i}(M,z)|^2}{\Omega_{\rm sky}} \frac{|\tilde{y}_{\ell_j}(M,z)|^2}{\Omega_{\rm sky}} + \cdots ,
\end{equation}
\end{widetext}
%
the term linear in~$\omega$ generates the mean of $C_\ell^{\rm tSZ, 1h}$ given by Eq.~\ref{eq:tsz_power_spectrum}, and the quadratic term yields the covariance $M_{\ell \ell'}$ of $C_\ell^{\rm tSZ, 1h}$ between different multipoles, given by
\begin{equation}
        M_{\ell\ell'} = \sigma_{\ell,G}^2
        \delta_{\ell \ell'} + \frac{T_{\ell\ell'}^{yy}}{4 \pi f_{\rm sky}},
\label{eq:tszcovmat_th}
\end{equation}
where the trispectrum $T_{\ell\ell'}^{yy}$ is given by
\begin{equation}
T_{\ell\ell'}^{yy} = \int\! dz \; \frac{dV}{dz\,d\Omega}
\int\! d\ln M \; \frac{dn}{d\ln M}
\left| \tilde{y}_{\ell} \right|^2  \left| \tilde{y}_{\ell'} \right|^2.
\label{eq:trispectrum}
\end{equation}
We work with binned bandpowers, whose covariance matrix is written as:
\begin{equation}
\begin{split}
        M_{bb'} =\;& \frac{2(C_{\ell_b}^{\rm tSZ,\,1h})^2}{(2\ell_b+1)f_{\rm sky}\,\Delta\ell_b}\,\delta_{bb'}\\
        &+\; \frac{\bar{T}_{bb'}^{yy}}{4\pi f_{\rm sky}}\,,
\end{split}
\label{eq:tszcovmat}
\end{equation}
where $\ell_b$ is the effective multipole of bin $b$, $\Delta\ell_b$ its width, and the bin-averaged trispectrum is
\begin{equation}
        \bar{T}_{bb'}^{yy} = \frac{1}{|\mathcal{B}_b|\,|\mathcal{B}_{b'}|}\,\sum_{\ell\in\mathcal{B}_b}\sum_{\ell'\in\mathcal{B}_{b'}}\! T_{\ell\ell'}^{yy}\,,
\label{eq:trisp_binned}
\end{equation}
with $\mathcal{B}_b$ the set of integer multipoles in bin $b$.


\subsection{Comparison to empirical sampling distributions}

\begin{figure*}
    \centering
    \includegraphics[width=0.95\linewidth]{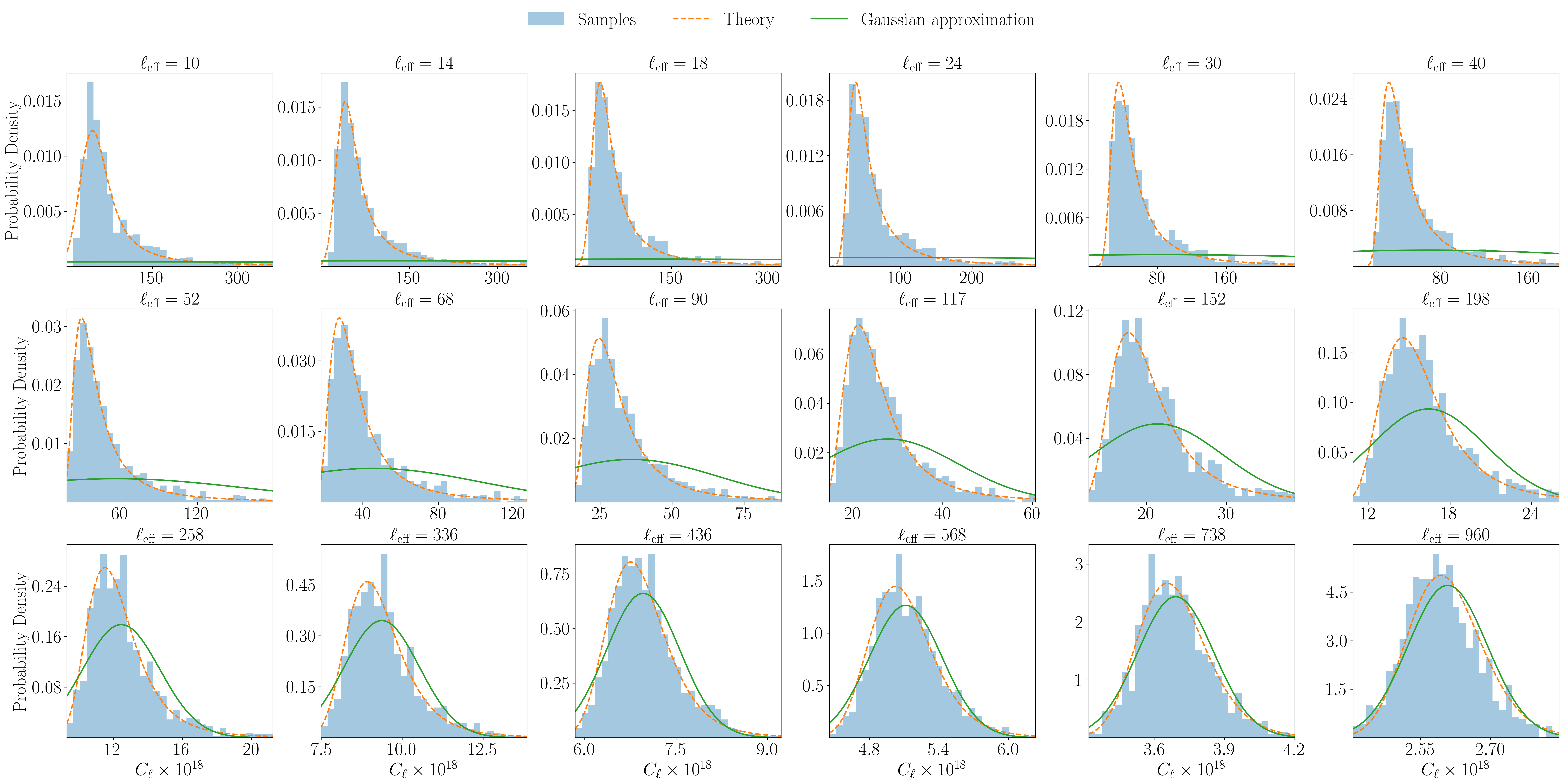}
    \caption{One-point distribution of the binned tSZ power spectrum from 1000 realizations (blue) compared with the theoretical prediction and Gaussian approximation. The theoretical prediction (orange dashed) is computed from Eq.~\eqref{1ptpdfourcase}; the Gaussian approximation (green solid) uses a binned version of the covariance given by Eq.~\eqref{eq:tszcovmat}. The distribution obtained from simulations agrees well with the theoretical prediction at all multipoles.
    }
    \label{fig:cl_distributions}
\end{figure*}

To quantify the non-Gaussianity of the tSZ power spectrum, we generate 1000 realizations of full-sky tSZ maps using halo-based simulations (described in Sec.~\ref{subsec:halobasedsims}) at the fiducial cosmology.
The power spectrum is computed from each realization, binned into 18 multipole bins. Figure~\ref{fig:cl_distributions} shows the distribution of these binned power spectra compared to the analytic prediction based on numerical evaluation of Eq.~\eqref{1ptpdfourcase} for the 1-point PDF. The agreement is excellent, reproducing the highly skewed distributions at low multipoles.
We also compare the empirical distribution of the 1000 realizations with a Gaussian approximation 
of the binned covariance matrix $M_{bb'}$. As can be seen, the Gaussian approximation (green curve) fits the simulated data poorly at low multipoles, where the power spectrum is dominated by a small number of massive clusters, but improves at higher multipoles as the number of contributing clusters increases. The Gaussian prediction shown here includes both terms in Eq.~\eqref{eq:tszcovmat}, with the trispectrum dominating in this case. The trend toward Gaussian distributions at higher multipoles reflects the underlying Gaussian limit of the Poisson distribution from the central limit theorem, not an increasing contribution from the $\sigma_{\ell,G}$ term, which only captures the scatter in the power spectrum at fixed number of clusters.


For cosmological inference, we require the full $n$-point PDF since power spectra at different multipoles are not independent.
The integral in Eq.~\eqref{nptjointpdf} can be computed numerically for low-dimensional cases. However, for high-dimensional joint distributions, the dimensionality of the integral increases dramatically, and the full analytic evaluation of the $n$-point joint PDF becomes intractable. In such cases, SBI methods can provide an alternative route that does not require an explicit evaluation of the likelihood. SBI methods can estimate the underlying likelihood and perform parameter inference using simulations, as we explore in Secs~\ref{sec:halobasedsims} and~\ref{sec:sbi}.

\section{Gaussian likelihood formalism}
\label{sec:lklformalism} 


Cosmological constraints can be obtained by modeling the observed tSZ power spectrum using Bayesian inference methods. In real data, the total power spectrum of Compton-$y$ maps includes the tSZ signal, in addition to residuals of astrophysical foregrounds and the instrumental noise. Here, we consider three astrophysical foreground components: the cosmic infrared background (CIB), shot noise from radio sources (RS) and from the infrared (IR) point sources that make up the CIB. These templates
represent the foreground power that remains after applying the component-
separation weights to multifrequency foreground models following the Planck
\(y\)-map power-spectrum analysis \citep{ade2013planck}. In addition, following the
Planck \(y\)-map power-spectrum analysis, we include an empirical
correlated-noise (CN) template \citep{Planck:2015vgm}. This CN contribution represents residual scale-dependent noise correlations in the
component-separated \(y\)-map spectra, which become important on small angular
scales. Hence, the model for the observed $y$-map power spectrum is

\begin{equation}
\begin{aligned}
C_{\ell}^{yy} \simeq\;& C_{\ell}^{\rm tSZ,\rm 1h}
+ A_{\rm CIB}\,C_{\ell}^{\rm CIB}
+ A_{\rm IR}\,C_{\ell}^{\rm IR} \\
&+ A_{\rm RS}\,C_{\ell}^{\rm RS}
+ A_{\rm CN}\,C_{\ell}^{\rm CN} \, ,
\end{aligned}
\label{tsz_base_model}
\end{equation}
where $C_{\ell}^{\rm CIB}$, $C_{\ell}^{\rm IR}$, $C_{\ell}^{\rm RS}$ and $C_{\ell}^{\rm CN}$ are templates of the foreground residuals and correlated noise terms, while the coefficients \(A_{\rm CIB}\), \(A_{\rm IR}\),
\(A_{\rm RS}\), and \(A_{\rm CN}\) are nuisance amplitudes. The templates are
normalized to the fiducial residual levels estimated in the Planck
\(y\)-map power-spectrum analysis. In practice, however, the residual foreground levels are uncertain and the different components are partially degenerate with each other and with the tSZ signal.



Traditional parameter inference methods for the tSZ power spectrum typically assume a Gaussian likelihood, given by $-2\ln \mathcal{L} = \chi^2+\ln|\mathbf{M}|+ {\rm const.}$, with
\begin{equation}
    \chi^2 = (\hat{\mathbf{C}}-\mathbf{C})^T\mathbf{M}^{-1}(\hat{\mathbf{C}}-\mathbf{C}),
\end{equation}
where $\hat{\mathbf{C}}$ and $\mathbf{C}$ are vectors containing the observed and theoretical tSZ power spectra $D_\ell = \ell(\ell+1)C_\ell/(2\pi)$ evaluated at all effective multipoles, and $\mathbf{M}$ is the fixed covariance matrix calculated using Eq.~\eqref{eq:tszcovmat} evaluated at fixed cosmology. The posterior distribution of the model parameters is then sampled within a Markov Chain Monte Carlo (MCMC) framework, typically using the Metropolis–Hastings (MH) algorithm.

However, as we have seen in Fig.~\ref{fig:cl_distributions}, this Gaussian likelihood assumption fails to capture the skewness of the tSZ power spectrum distribution at low multipoles.
A traditional Gaussian likelihood analysis may not provide an accurate description of the distribution of the data and may potentially lead to misestimated parameter constraints. This motivates considering alternative methods beyond the Gaussian likelihood analysis. 

To evaluate the theoretical tSZ power spectrum and the covariance matrix at each point in parameter space, we use our Python package \texttt{tszsbi}\footnote{\url{https://github.com/licongxu/tszsbi}}. It computes the 1-halo tSZ power spectrum and the trispectrum contribution to the covariance using \texttt{CosmoPower} emulators \citep{Spurio_Mancini_2022, bolliet2023highaccuracyemulatorsobservableslambdacdm} that are used to build \texttt{class\_sz} \citep{Bolliet:2023eob,Bolliet:2025oqo}. Written in JAX \citep{jax2018github}, \texttt{tszsbi} supports automatic differentiation and Just-In-Time compilation for fast likelihood evaluation.

\section{Forward modeling}
\label{sec:halobasedsims}


Simulation-based inference allows us to infer model parameters using forward-modeled simulations without sampling from an explicit likelihood. To ensure the parameter constraints are unbiased, a robust forward-modeling pipeline must be constructed to produce realistic mock observations across the parameter space of interest. In this section, we will first introduce a Gaussian likelihood-based simulator, which serves as the first validation of our SBI pipeline. We then move to a more realistic halo-based forward model, which captures the non-Gaussian statistics of the tSZ power spectrum.

\subsection{Forward modeling for the Gaussian likelihood}
\label{subsec:lklbasedsbi}
As a first demonstration, we formulate the Gaussian likelihood analysis described in Sec.~\ref{sec:lklformalism} in an SBI-equivalent setting. We begin by constructing a forward simulator to simulate $C_\ell^{yy}$ for given input of cosmological and foreground parameters. We assume the probability distribution of the tSZ power spectrum follows a multivariate Gaussian distribution. The observational uncertainties can be modeled by adding Gaussian noise using Cholesky decomposition of the fixed covariance matrix, i.e., $\mathbf{M} = \mathbf{LL}^T$, where $\mathbf{L}$ is a lower-triangular matrix. The total simulated power spectrum is given by $C_{\ell}^{yy, \rm sim}= C_{\ell}^{yy, \rm signal}+ N_{\ell}$, where $C_{\ell}^{yy, \rm signal}$ is given by Eq.~\eqref{tsz_base_model}, and $\mathbf{N} = \mathbf{L}\mathbf{v}$, where $\mathbf{v}$ is a vector with elements drawn from a standard normal distribution. This method enables us to generate parameter-simulation pairs consistent with the Gaussian likelihood analysis, to which SBI can then be applied.


\subsection{Forward modeling using halo-based approach}
\label{subsec:halobasedsims}
To capture realistically the non-Gaussianity of the tSZ power spectrum due to massive and local clusters, we adopt a halo-based forward-modeling pipeline. In this pipeline, we first generate mock cluster catalogs for a given set of cosmological parameters, and then paint the clusters onto sky maps at random positions using their corresponding pressure profiles. Mock halo catalogs are generated by Poisson sampling of halo masses and redshifts from the HMF of Tinker \textit{et al.}~\citep{Tinker:2008ff} using the \texttt{cosmocnc}\footnote{\url{https://github.com/inigozubeldia/cosmocnc}} package \citep{Zubeldia:2024lke}. The mass and redshift ranges of the halos are chosen to be $M \in [10^{14}, 10^{16}]M_\odot$, and $z \in [0.005, 3]$. Note that we adopt a higher minimum mass threshold than that used in the likelihood analysis of the real \textit{Planck} data in \citet{Bolliet:2017lha}, as the large-scale ($\ell < 10^3$) tSZ power is dominated by massive clusters, and the power spectrum converges rapidly for this $M_{\rm min}$, as indicated in Sec.~\ref{tszmzdep}. This also significantly reduces the computational cost in the subsequent map-generation step, since the computational time of generating sky maps from halo catalogs scales with the number of clusters as $O(N_{\rm clusters})$. Given the steep low-mass slope of the HMF, excluding low-mass halos substantially reduces the number of clusters in the catalogs while causing only a slight loss of tSZ power at high multipoles, where the signal is already dominated by foreground residuals and instrumental noise. 


The HMF is evaluated on a grid of $n_M = 5000$ mass points and $n_z = 5000$ redshift points. The total number of clusters in each catalog is then drawn from a Poisson distribution with mean given by the integral of the HMF over the chosen mass and redshift ranges. The $(M, z)$ pairs of clusters in each catalog are subsequently sampled from the continuous distribution given by the HMF. Halo positions are sampled uniformly across the sky.  We validate our synthetic catalogs by comparing the mean number counts across mass and redshift using 100 synthetic catalogs at fixed cosmology and the theoretical prediction of the input HMF from \texttt{cosmocnc}, as indicated in Appendix \ref{app:gencatprecision}. No significant biases are found.



\begin{figure*}
    \centering
    \includegraphics[width=0.99\linewidth]{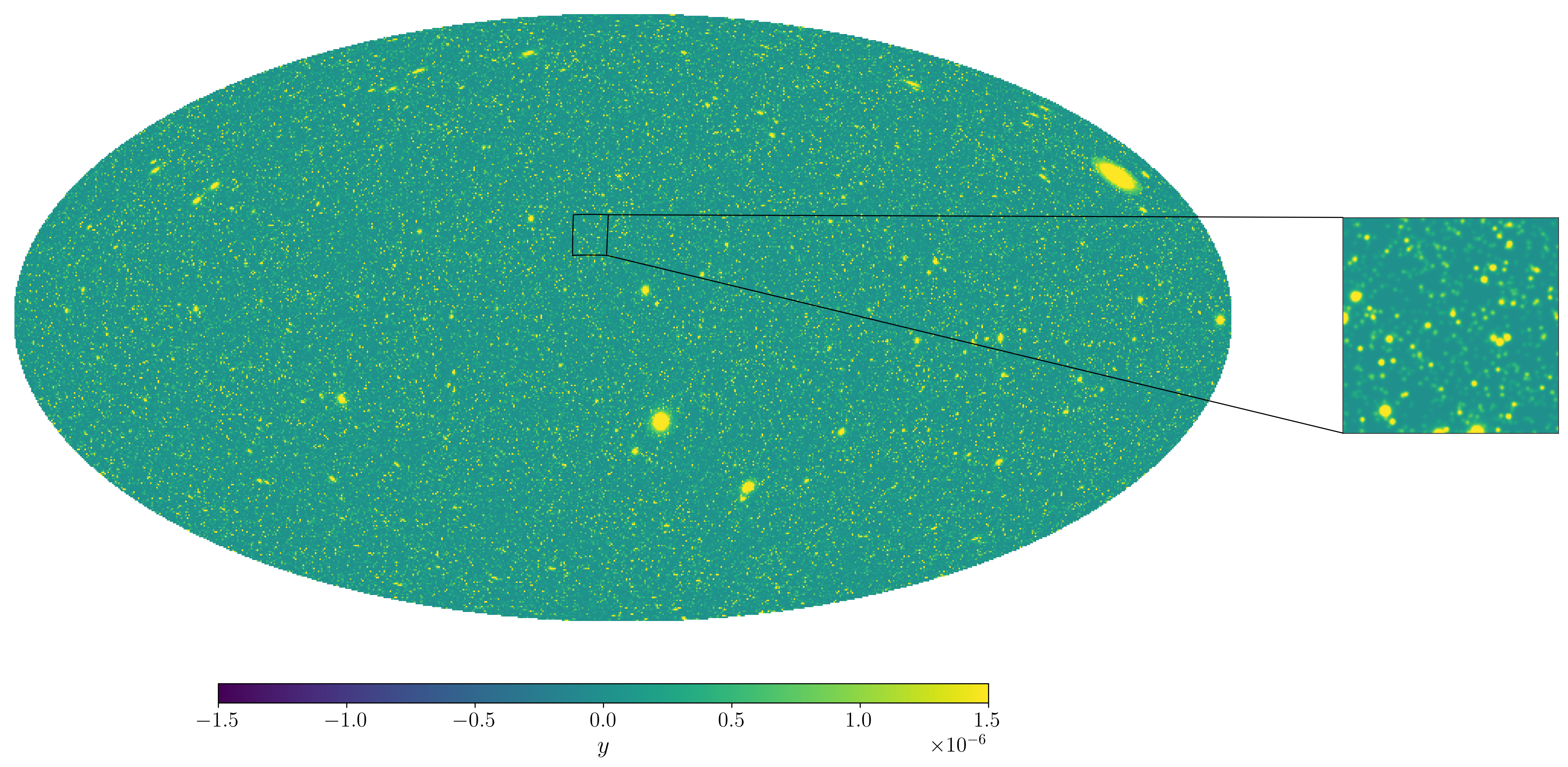}
    \caption{Example of a noiseless full-sky Compton-$y$ map and a $10^{\circ} \times 10^{\circ}$ patch extracted from it. The maps are generated using a modified version of \texttt{XGPaint} with $\theta_{\mathrm{FWHM}} = 10$~arcmin and $N_{\mathrm{side}} = 1024$.}
    \label{fig:fullskyvs10deg}
\end{figure*}

The full-sky maps are generated with a modified version of the \texttt{XGPaint} package\footnote{\url{https://github.com/WebSky-CITA/XGPaint.jl}}. Each halo in the catalog is assigned the~\citetalias{Arnaud_2010} pressure profile and convolved with a Gaussian beam with $\theta_{\text{FWHM}}=10\,\text{arcmin}$. The pressure profile is truncated at an angular radius equal to the larger of $4\theta_{500}$ or $2\theta_{\rm FWHM}$, where $\theta_{500} = r_{500c}/d_A$.
The maps are constructed on \texttt{Healpix.jl} \citep{2021ascl.soft09028T} grids with resolution $N_{\rm side}=1024$. An example of the resulting noiseless full-sky Compton-$y$ map and a $10^{\circ} \times 10^{\circ}$ patch extracted from it are shown in Fig.~\ref{fig:fullskyvs10deg}. 
Foreground residuals are added separately as Gaussian random fields generated from their residual beam-convolved power spectra. The angular power spectra are computed from these maps using the \textit{Planck} 2015 binning scheme, with bandpowers defined at the 18 effective multipoles $\ell_{\mathrm{eff}}$ of Ref.~\citep{Planck:2015vgm}. Within each bin, we apply uniform weights to $D_\ell$ when averaging over multipoles. 


\begin{figure}[b]
    \centering

    \includegraphics[width=0.95\linewidth]{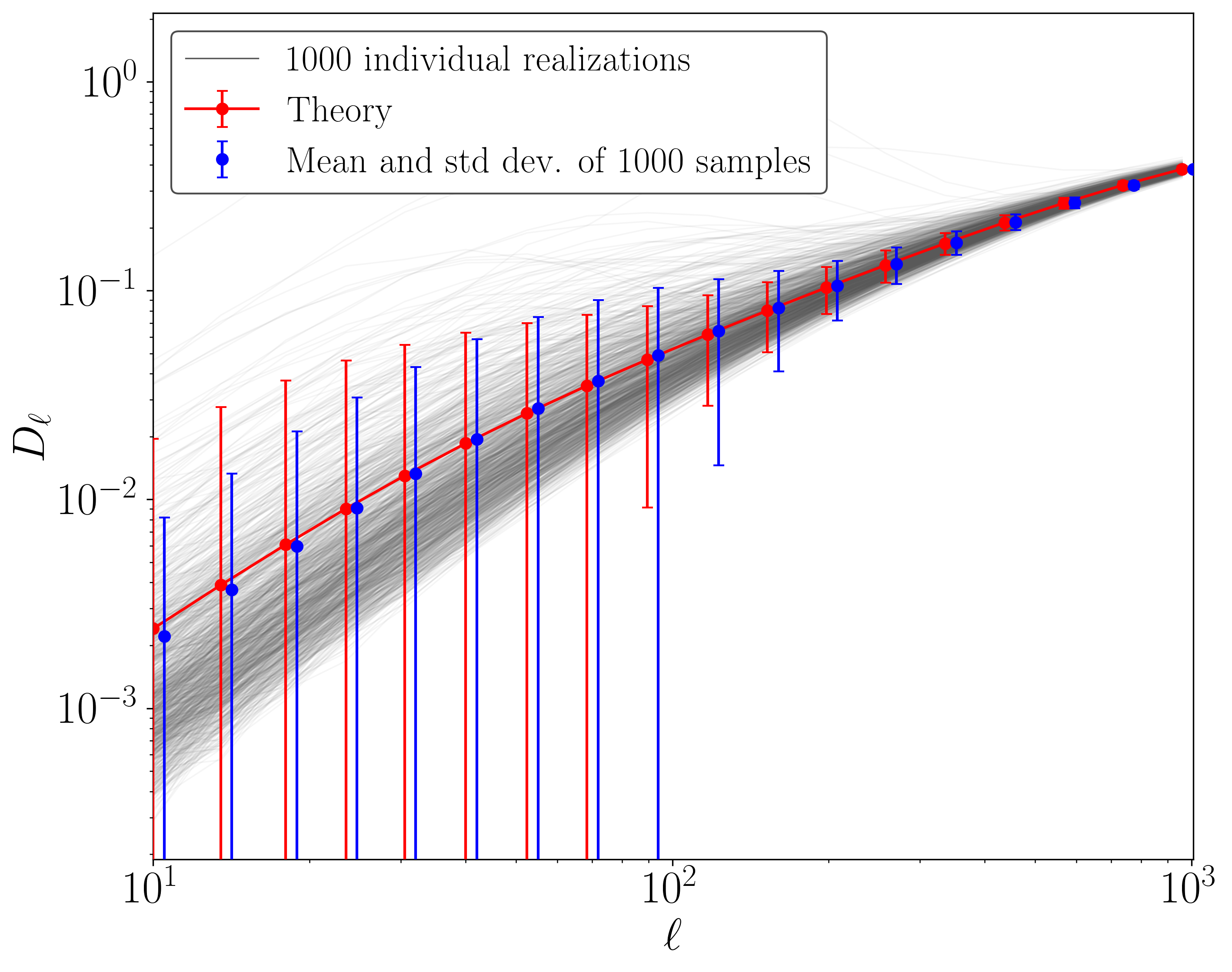}

    \vspace{0.5em}

    \includegraphics[width=0.95\linewidth]{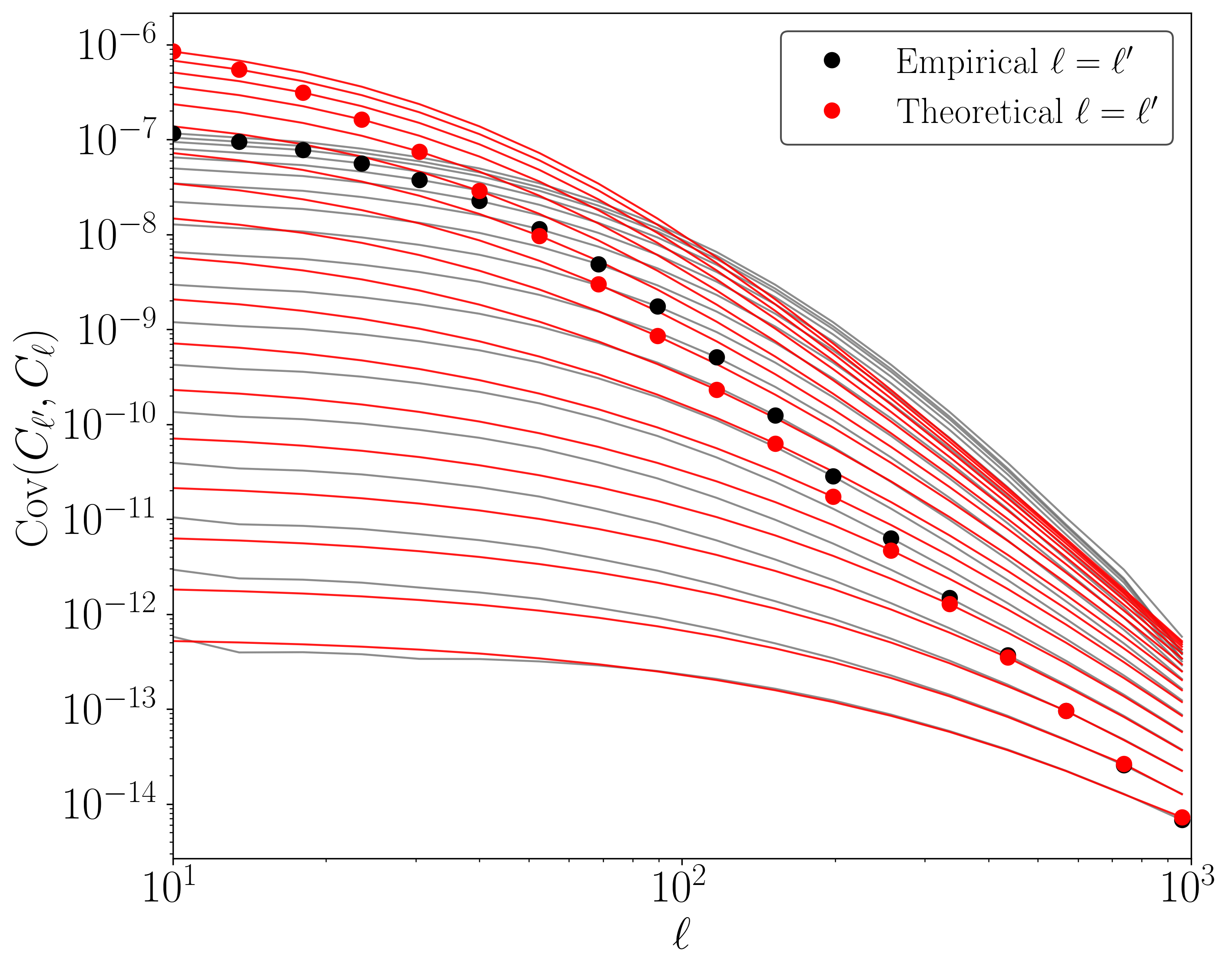}

    \caption{\textit{Upper}: Mean and standard deviation of the tSZ power spectrum
    computed from 1000 simulated maps (blue) compared with the corresponding theoretical
    predictions (red). The power spectrum of each individual realization is shown in gray. The empirical standard deviation agrees well with the theoretical error bars at high $\ell$, while larger fluctuations are seen at low $\ell$, where the variance is less well estimated from a finite number of simulations.
    \textit{Lower}: Covariance matrix estimated from 1000 simulations (gray) and the
    theoretical prediction (red). Each line represents a different multipole $\ell'$ in the range $10$--$1000$ (top to bottom). Filled circles indicate the diagonal elements of the
    covariance matrix. The mean simulated tSZ power spectrum agrees with the theoretical
    prediction overall, while showing larger scatter at low multipoles, as expected from
    the stronger sample variance in this regime.}
    \label{fig:tszps_combined}
\end{figure}

To validate the sky maps generated, we produce 1000 realizations at the fiducial cosmology and hydrostatic bias parameter, and compute the mean and covariance of the resulting tSZ power spectra. 
Figure~\ref{fig:tszps_combined} shows the mean and standard deviation of the tSZ power spectrum computed from 1000 simulated maps, as well as the covariance of $C_\ell$ between multipoles. The simulated mean spectrum is consistent with the theoretical prediction to within a few percent over the full multipole range, with subpercent-level agreement achieved at high multipoles, where the variance is small and most of the constraining power is concentrated. 
The covariance matrix obtained from simulations agrees well with the theoretical prediction at high multipoles. The agreement is poorer at low multipoles. This is because the variance of the error in the empirical covariance due to the finite number of simulations depends on the fourth connected moment of the sampling distribution of the $C_\ell$. At low multipoles, the fourth connected moment is much larger than the square of the variance due to the heavy tails (see Fig.~\ref{fig:cl_distributions}), giving large scatter in our empirical estimates.


\section{SBI Implementation}
\label{sec:sbi}
We use the \texttt{sbi} package \citep{BoeltsDeistler_sbi_2025}, which implements SBI algorithms using neural networks built in PyTorch. In this paper, we consider both NPE \citep{papamakarios2018fastepsilonfreeinferencesimulation} and NLE \citep{papamakarios2019sequentialneurallikelihoodfast} to infer the posterior distribution. NPE models the posterior directly from parameter--data pairs, while NLE learns the likelihood, requiring an additional sampling step to obtain the posterior. In both cases, the neural networks are trained on a set of simulated data and corresponding cosmological and nuisance parameters, allowing efficient inference on the observed data.


We generate 9000 simulations for the likelihood-based SBI approach described in Sec.~\ref{subsec:lklbasedsbi}, and 18,000 simulations for the halo-based SBI method using our pipeline described in Sec.~\ref{subsec:halobasedsims}. Each SBI model is trained using an 80:20 train–validation split and optimized with the AdamW optimizer \citep{loshchilov2019decoupledweightdecayregularization}. To stabilize training and prevent overfitting, we apply gradient clipping with a maximum norm of 5.0 and implement early stopping, terminating training if no improvement in validation loss is observed after 50 epochs. We adopt a Masked Autoregressive Flow (MAF) architecture to model the posterior or likelihood distribution \citep{papamakarios2018maskedautoregressiveflowdensity}. The number of transformation layers and the size of the hidden layers are treated as hyperparameters and optimized with respect to the best validation loss using \texttt{wandb} \citep{wandb}. The hyperparameters used are summarized in Table~\ref{tab:hyperparams}.


We also find that training a single network repeatedly can lead to noticeable fluctuations in the inferred posterior, even when the same architecture and hyperparameters are used \citep{alvey2025simulationbasedinferencedeepensembles}. This can be due to the stochastic elements in training, such as random initialization and optimization paths. To mitigate instabilities arising from stochasticity in neural network training, we employ an ensemble of 10 independently trained networks and form a uniformly weighted mixture of their posteriors or likelihoods. By averaging the posteriors predicted by the ensemble, we obtain a more stable and robust estimate of the inferred posterior distribution.

\begin{table*}[ht]
\caption{Training hyperparameters for the Masked Autoregressive Flow (MAF) architecture used in the likelihood-based and halo-based SBI approaches. The table lists the neural network and optimization settings for both Neural Posterior Estimation (NPE) and Neural Likelihood Estimation (NLE) approaches.}
\label{tab:hyperparams}
\begin{ruledtabular}
\begin{tabular}{lcccc}
Hyperparameter & NPE (likelihood-based) & NLE (likelihood-based) & NPE (halo-based) & NLE (halo-based) \\
\hline
Hidden features        & 80   & 20   & 150  & 30 \\
Number of transforms    & 15   & 6    & 10   & 18 \\
Number of components    & 5    & 5    & 5    & 5 \\
Number of bins          & 3    & 3    & 3    & 3 \\
Dropout probability     & 0.05 & 0    & 0.05 & 0 \\
Activation function     & ReLU & ReLU & $\tanh$ & $\tanh$ \\
Batch size              & 64   & 128  & 256  & 256 \\
Initial learning rate   & $2.382\times10^{-4}$ & $5.593\times10^{-3}$ & $4.871\times10^{-4}$ & $1.758\times10^{-3}$ \\
Learning rate decay factor ($\gamma$) & 0.9528 & 0.8304 & 0.8551 & 0.7341 \\
Scheduler step size     & 43 epochs & 22 epochs & 63 epochs & 36 epochs \\
\end{tabular}
\end{ruledtabular}
\end{table*}


\begin{table}[ht]
\caption{Priors imposed on cosmological and foreground parameters in likelihood analysis. We adopt uniform priors on all parameters, where $\mathcal{U}(a,b)$ denotes a uniform distribution between $a$ and $b$.}
\label{tab:prior}
    \centering
    \begin{ruledtabular}       
    \begin{tabular}{lc}
        Parameter             & Prior Distribution \\
        \hline
        $\Omega_b h^2$        & $\mathcal{U}(0.02, 0.025)$ \\
        $\Omega_c h^2$        & $\mathcal{U}(0.11, 0.13)$ \\
        $H_0$ (km\,s$^{-1}$\,Mpc$^{-1}$) & $\mathcal{U}(55, 90)$ \\
        $\ln(10^{10}A_s)$     & $\mathcal{U}(2.5, 3.5)$ \\
        $n_s$                 & $\mathcal{U}(0.94, 1.0)$ \\
        $B$                   & $\mathcal{U}(1.0, 2.0)$ \\
        $A_{\text{CIB}}$      & $\mathcal{U}(0, 5)$ \\
        $A_{\text{IR}}$       & $\mathcal{U}(0, 5)$ \\
        $A_{\text{RS}}$       & $\mathcal{U}(0, 5)$ \\
    \end{tabular}
    \end{ruledtabular}
\end{table}

\section{Results}
\label{sec:results}
In this section, we present the results of our parameter constraints using both the likelihood-based and halo-based SBI approaches. We first compute the parameter constraints by sampling from the Gaussian likelihood using \texttt{Cobaya} \citep{Torrado_Lewis_2021_Cobaya, cobaya_ascl} with the MH algorithm, and compare with those obtained from its SBI-equivalent settings to validate the SBI application in this task. We then compare the results obtained by halo-based SBI with those from the Gaussian likelihood to assess how the inferred parameter constraints change when the Gaussian-likelihood assumption is replaced by a halo-based forward model.


\subsection{Data}
In this work, we use two distinct data setups. For the likelihood-based analysis, we use the real power spectrum data measured by \textit{Planck}, with multipole bins ranging from $10 \leq \ell \leq 959.5$ following \citet{Bolliet:2017lha}. The likelihood analysis is performed directly on these data using the Gaussian likelihood described in Sec.~\ref{sec:lklformalism}, while the corresponding SBI comparison is trained on samples drawn from the same Gaussian likelihood with the noise drawn from the Cholesky decomposition of the covariance matrix (Sec.~\ref{subsec:lklbasedsbi}), so that it provides an SBI equivalent of the likelihood-based setup. By contrast, for the halo-based analysis, we instead define a mock observed power spectrum from a single realization of the halo-based forward model at the fiducial parameters as described in Sec.~\ref{subsec:halobasedsims}, and then compare halo-based SBI constraints against those obtained by applying the Gaussian-likelihood analysis to this same mock dataset. We neglect instrumental noise ($A_{\rm CN}=0$) in this setup. These steps allow us first to validate SBI against the standard likelihood-based approach on real data, and then assess how the inferred constraints change when the Gaussian-likelihood assumption is replaced by a halo-based forward model in a controlled mock-data case. In both setups, we vary the five cosmological parameters, the hydrostatic bias parameter $B$, and the three residual foreground amplitudes with the priors shown in Table~\ref{tab:prior}.

\subsection{Constraints from likelihood-based analysis}
\label{subsec:lklsbi}
\begin{figure}[b]
\includegraphics[width=0.95\linewidth]{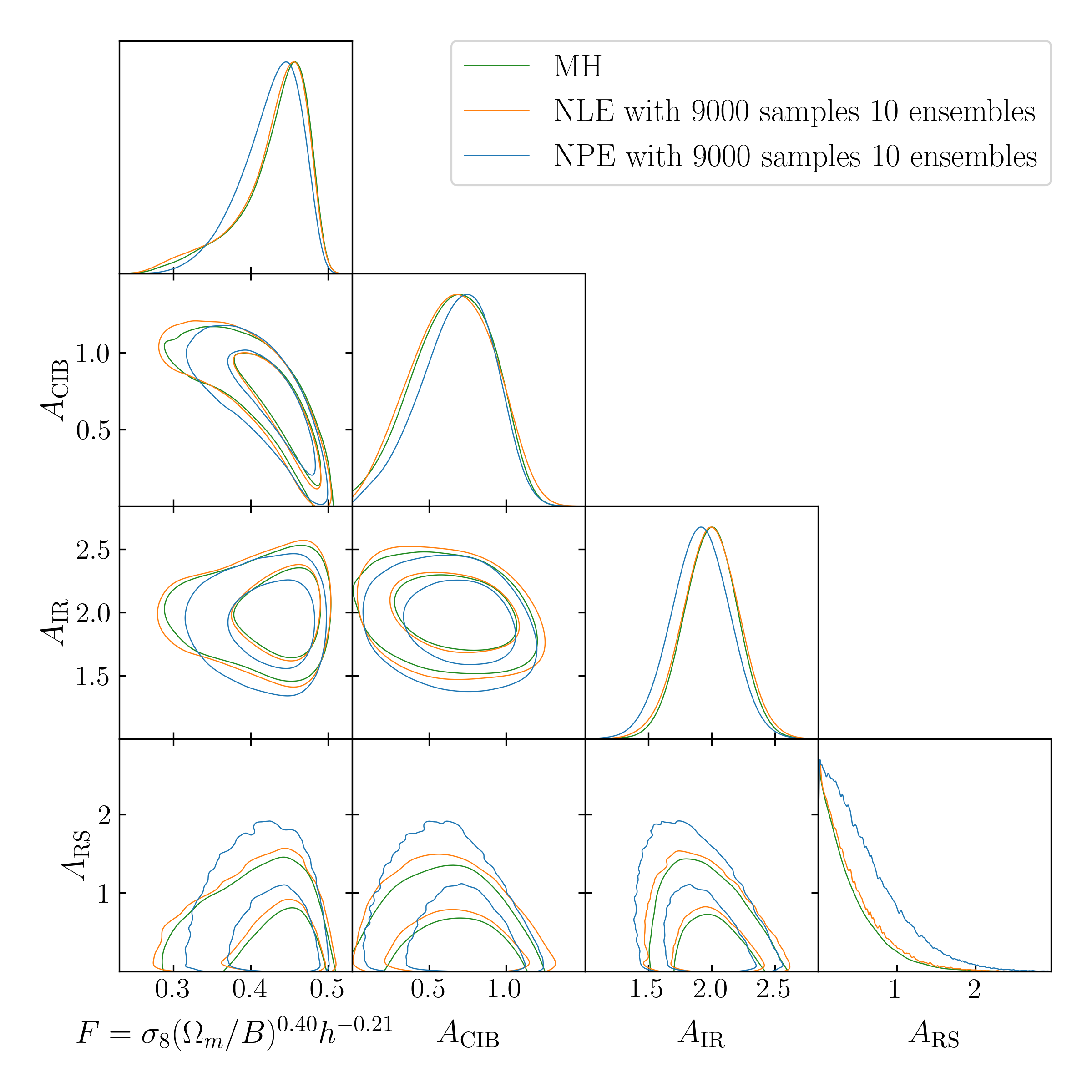}
\caption{\label{fig:posterior_lkl}
Posterior constraints from the Gaussian likelihood analysis and the SBI analyses applied to the real \textit{Planck} data measured by \citet{Bolliet:2017lha}. The green
contours show the Gaussian likelihood result, while the orange and blue
contours show the NLE and NPE results, respectively. Both SBI analyses are
trained using synthetic data generated with the same Gaussian power-spectrum
forward model assumed in the likelihood analysis. Comparing posterior means of each parameter, the NLE results agree with the likelihood results to
within \(0.05\sigma\), while the NPE results agree to within \(0.4\sigma\),
with the largest mean shift occurring for \(A_{\rm IR}\). Contours indicate
the 68\% and 95\% credible regions.
}
\end{figure}

Figure~\ref{fig:posterior_lkl} compares the posterior constraints obtained from the Gaussian-likelihood analysis using the MH sampler with those inferred using SBI methods under the same Gaussian likelihood assumption. For the MH reference analysis, convergence is assessed using the Gelman–Rubin criterion $\hat{R}-1<0.01$ \citep{GelmanRubin1992}. The corresponding parameter constraints are summarized in Table~\ref{tab:parameter_constraints_lkl}. We find excellent agreement between the NLE posterior and the MH reference posterior, with the differences in the mean inferred values being less than $0.05\sigma$. The NPE posteriors show poorer agreement with the likelihood analysis, with the largest deviation being about $0.4\sigma$ for $A_{\rm IR}$. This likely reflects the intrinsic difficulty for NPE to learn the highly non-Gaussian posterior with a MAF neural network architecture. This difficulty may arise from degeneracies between the tSZ power spectrum signal and the foreground templates, which make the posterior highly non-Gaussian. The positivity prior on $A_{\rm RS}$ may also contribute by truncating part of the posterior support, although in principle the MAF model should be able to learn such prior-induced boundaries. We note that the contribution from radio sources is poorly constrained compared to the cosmological constraint $F$ and other foreground residual amplitudes. This is primarily due to degeneracies between the radio-source template and the other foreground components and the tSZ signal, as also noted in \citet{Planck:2015vgm}. Nevertheless, we find excellent agreement between the NLE and MH posteriors, despite the template degeneracies leading to numerical difficulties. This demonstrates the ability of SBI to produce reliable posterior estimates for this density estimation task.




\begin{table}[ht]
\caption{Comparison of parameter constraints from the Gaussian likelihood analysis and NPE and NLE SBI-equivalent analyses using real \textit{Planck} data measured by \citet{Bolliet:2017lha}. The posterior distributions are shown in Fig.~\ref{fig:posterior_lkl}. The quoted parameter constraints correspond to 68\% credible bounds.}
\label{tab:parameter_constraints_lkl}
\renewcommand{\arraystretch}{1.5}
\centering
\begin{ruledtabular}
\begin{tabular}{lccc}
Parameter & MH & NLE & NPE \\
\colrule
$F$ & $0.430^{+0.054}_{-0.018}$ & $0.428^{+0.055}_{-0.018}$ & $0.425^{+0.049}_{-0.025}$ \\
$A_{\mathrm{CIB}}$ & $0.66^{+0.30}_{-0.23}$ & $0.65^{+0.30}_{-0.25}$ & $0.68^{+0.28}_{-0.20}$ \\
$A_{\mathrm{IR}}$ & $1.99 \pm 0.21$ & $1.99 \pm 0.22$ & $1.91 \pm 0.23$ \\
$A_{\mathrm{RS}}$ & $0.39^{+0.073}_{-0.39}$ & $0.422^{+0.090}_{-0.42}$ & $0.58^{+0.13}_{-0.57}$ \\
\end{tabular}
\end{ruledtabular}
\end{table}

\subsection{Constraints from halo-based SBI}
\label{subsec:halosbi}
We next turn to the halo-based forward-modeling setup, with a simple scenario in which instrumental noise is neglected. The data vector consists of the tSZ power spectra computed from maps generated from the simulated halo catalogs, to which we add residual foregrounds (CIB, IR, and RS) drawn as Gaussian random fields. For benchmarking purposes, we take as our data a single realization of the tSZ map drawn from the 1000 simulations described in Sec.~\ref{subsec:halobasedsims}, combined with one realization of the foreground residuals generated with parameters $A_{\mathrm{CIB}} = 0.66$, $A_{\mathrm{IR}} = 2.04$, and $A_{\mathrm{RS}} = 0.0004$, which are chosen to be consistent with the foreground-amplitude constraints obtained from the \textit{Planck} analysis in Sec.~\ref{subsec:lklsbi}.

\begin{figure}[b]
\includegraphics[width=0.95\linewidth]{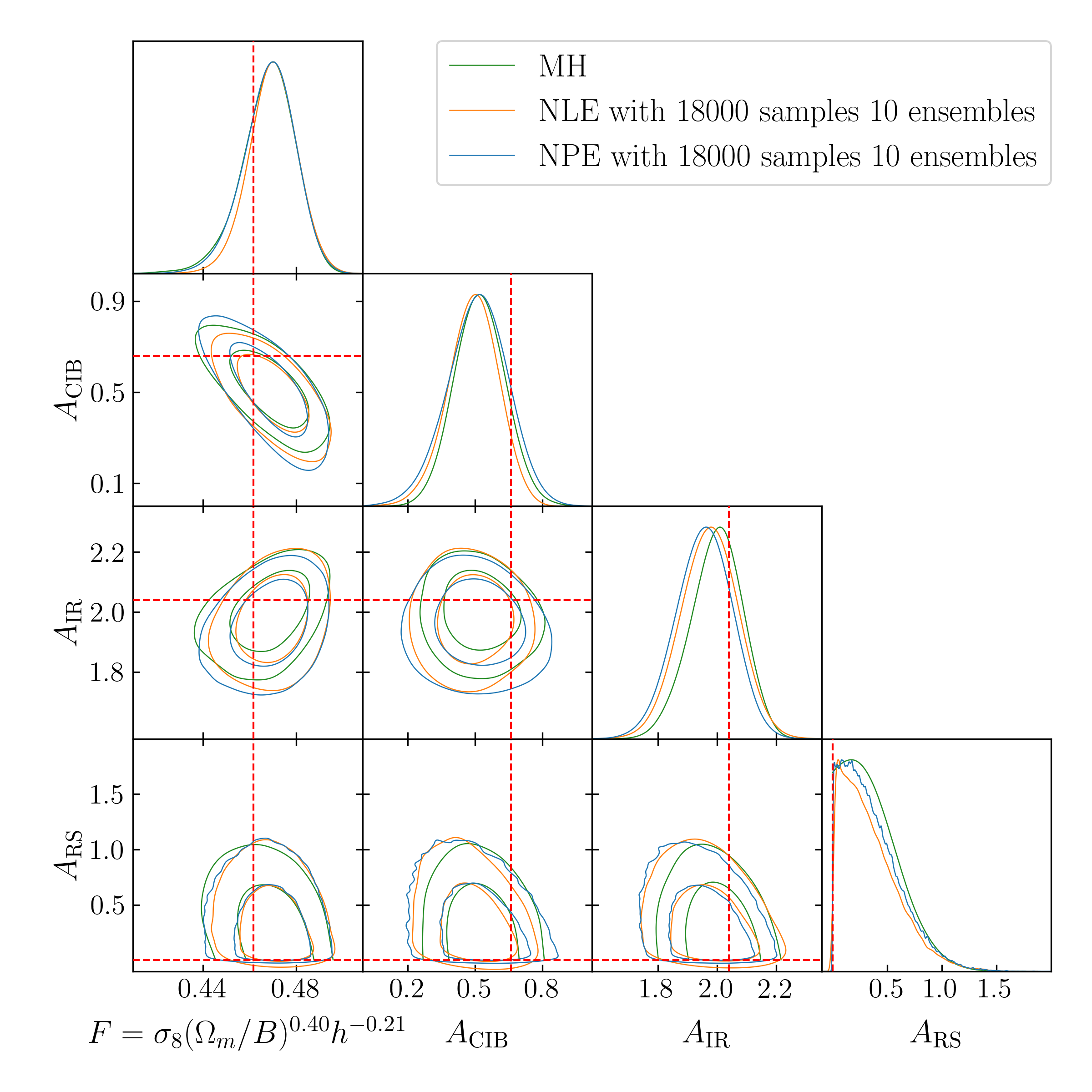}
\caption{\label{fig:posterior_paint} Parameter constraints from a simulated halo-based realization of the tSZ power spectrum. Green contours are for an approximate Gaussian-likelihood analysis while SBI results are shown for NLE (orange) and NPE (blue).
Red dashed lines indicate the true parameter values used to generate the benchmark simulation. Contours indicate the 68\% and 95\% credible regions.}
\end{figure}

Figure~\ref{fig:posterior_paint} shows the posterior contours obtained from both SBI approaches and the (approximate) Gaussian likelihood. The corresponding parameter constraints are summarized in Table~\ref{tab:parameter_constraints_paint}. All three methods recover constraints that are consistent with the input parameter values.
Remarkably, both the NLE and NPE approaches yield the posterior distributions for the cosmological parameter $F$ in excellent agreement with the likelihood-based result, despite the fact that the likelihood is highly non-Gaussian at low $\ell$. However, there are some differences in the foreground residual amplitude constraints from SBI and the likelihood-based method. We note that the SBI-based methods yield slightly larger uncertainties in all three amplitudes compared to the likelihood analysis. The constraint on $A_{\rm CIB}$ from NLE is $0.3\sigma$ lower than the likelihood-based result, while the $A_{\rm IR}$ value obtained from NPE is $0.5\sigma$ lower than the likelihood-based constraint. For $A_{\rm RS}$, both SBI methods yield the inferred means below the mean inferred from the likelihood-based analysis. Based on the benchmark on the real \textit{Planck} data in Sec.~\ref{subsec:lklsbi}, where NLE showed closer agreement with the MH reference posterior, we regard the NLE posterior as the more robust SBI estimate in this test. This is further supported by the validation test in Sec.~\ref{subsec:validation_sbi}, which shows that NLE can accurately learn the non-Gaussian likelihood structure across the multipole range.


Overall, despite the true likelihood being non-Gaussian, the SBI-based constraints on the cosmology-derived parameter $F$ are in excellent agreement with those obtained from the Gaussian likelihood analysis. The discrepancies in the foreground residual constraints may arise from several factors. For example, 
\begin{itemize}
    \item the true likelihood certainly differs from the Gaussian approximation due to the intrinsic non-Gaussianity of the signal;
    \item the SBI-related artifacts, including incomplete convergence due to a finite number of simulations and stochasticity in neural network training; and
    \item mismatches between the simulations and the theoretical model, since the SBI posterior is learned entirely from \texttt{XGPaint} simulations, while the likelihood analysis is based on theory and applied to one realization generated by the same simulation pipeline.
\end{itemize}  

The discrepancies are most likely driven by the first factor, since the power spectrum remains slightly skewed even at high multipoles. SBI-related artifacts are less likely given our benchmark experiment in Sec.~\ref{subsec:lklsbi}, which shows good agreement between SBI and the likelihood-based analysis for the same kind of data under the Gaussian-likelihood assumption. Nevertheless, we carry out further validation tests to assess possible SBI artifacts and simulation biases in Sec.~\ref{subsec:validation_sbi}. 

\begin{table}[ht]
\caption{Comparison of parameter constraints from a Gaussian likelihood (MH) and SBI with NLE and NPE applied to a simulated halo-based realization of the tSZ power spectrum.
The posterior distributions are shown in Fig.~\ref{fig:posterior_paint}.
All quoted parameter constraints correspond to 68\% credible bounds.}
\label{tab:parameter_constraints_paint}
\renewcommand{\arraystretch}{1.5}
\centering
\begin{ruledtabular}
\begin{tabular}{lccc}
Parameter & MH & NLE & NPE \\
\colrule
$F$ & $0.468^{+0.013}_{-0.0097}$ & $0.470 \pm 0.010$ & $0.468^{+0.012}_{-0.0098}$ \\
$A_{\mathrm{CIB}}$ & $0.52 \pm 0.11$ & $0.49^{+0.12}_{-0.11}$ & $0.51^{+0.14}_{-0.13}$ \\
$A_{\mathrm{IR}}$ & $1.998^{+0.094}_{-0.081}$ & $1.974 \pm 0.096$ & $1.955 \pm 0.095$ \\
$A_{\mathrm{RS}}$ & $0.37^{+0.094}_{-0.37}$ & $0.35^{+0.10}_{-0.33}$ & $0.358^{+0.094}_{-0.35}$ \\
\end{tabular}
\end{ruledtabular}
\end{table}

\subsection{Reconstructed tSZ power spectrum}
\begin{figure}
    \centering
    \includegraphics[width=0.99\linewidth]{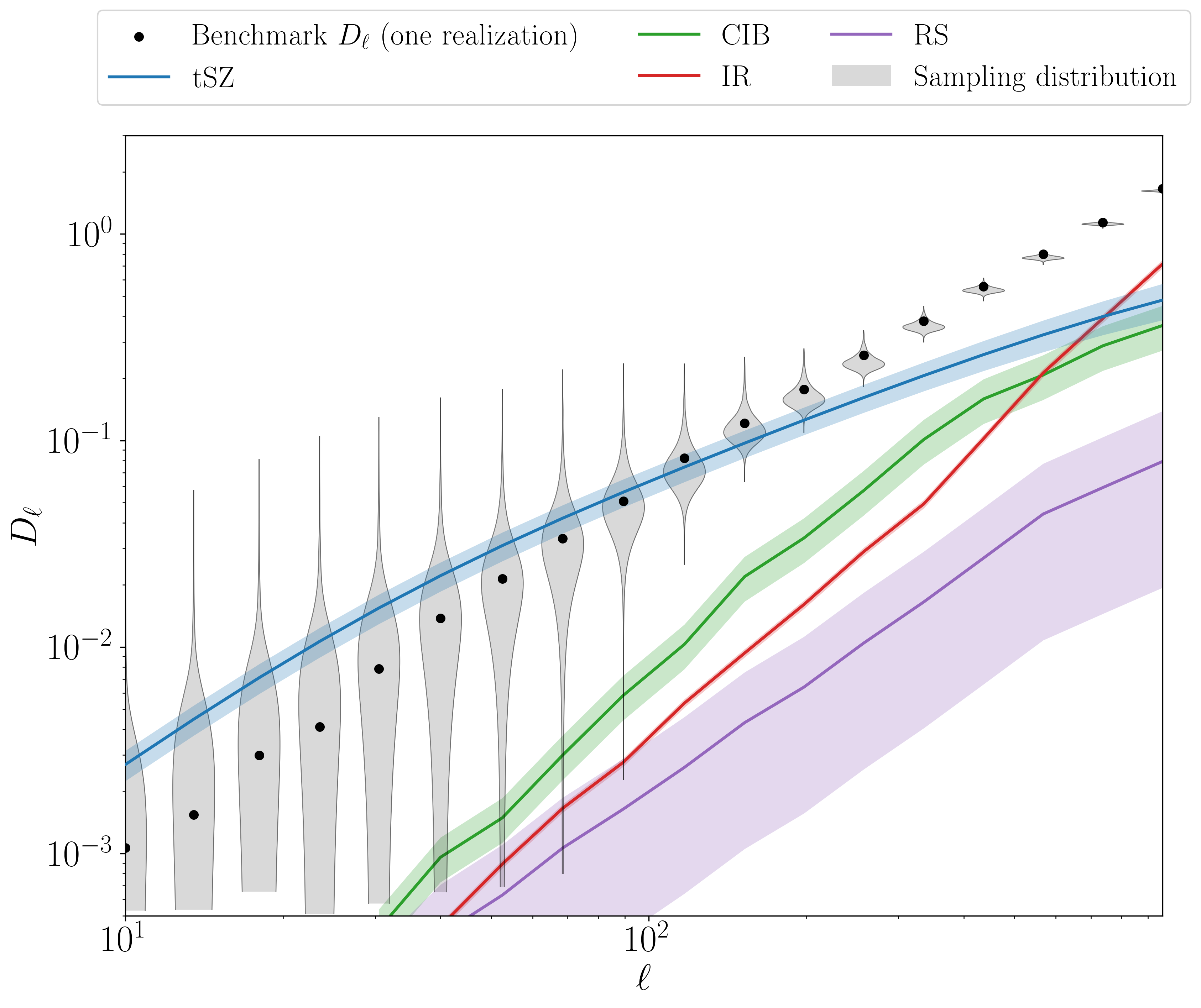}
    \caption{Reconstructed tSZ power spectrum and foreground residuals for a single simulated realization using the NLE method. The black dots represent the total binned power spectrum computed from one realization of simulated maps with tSZ, CIB, IR and RS included. The gray violin plots show the sampling distribution of the 1000 simulated bandpowers in each multipole bin, illustrating the positively skewed distribution of the power spectrum, especially at low multipoles. The blue curve and shaded band show the mean and $1\sigma$ uncertainty of the reconstructed tSZ spectrum obtained from 1000 posterior samples. Foreground residuals are shown as template-shaped curves with correlated amplitude uncertainties translated into smooth shaded bands. The uncertainties are represented as smooth shaded bands, because all amplitude uncertainties (for both tSZ and foregrounds) create fully correlated fluctuations across multipoles.}
    \label{fig:tszps_reconstructed}
\end{figure}

Figure~\ref{fig:tszps_reconstructed} shows the reconstructed power spectra of the tSZ signal and foreground residuals from the halo-based setup in Sec.~\ref{subsec:halosbi} for the NLE method.
The black dots denote the total power spectrum of the mock observation, while the gray violin plots show the sampling distribution of the 1000 simulated bandpowers in each multipole bin. This ensemble of simulated bandpowers is generated at fixed cosmological and residual foreground parameters, matching the mock observation.
To obtain the reconstructed tSZ signal, we draw 1000 samples from the posterior distribution of the parameters and compute, at each multipole, the theoretical 1-halo tSZ power spectrum for each sample. 
Since we use fixed templates for foreground residuals, the mean and standard deviation of each foreground amplitude translate directly into the mean and uncertainty band of the corresponding foreground residual power spectrum. Because all amplitude uncertainties (for both tSZ and foregrounds) are fully correlated across multipoles, we represent the reconstructed uncertainties using smooth shaded bands rather than independent error bars. The reconstructed tSZ spectrum is detected with a mean signal-to-noise ratio of 5.6. This is consistent with the posterior constraint on $F$ in Table~\ref{tab:parameter_constraints_paint} with a fractional uncertainty of around 2.4\%. Since $F$ controls the overall amplitude of the tSZ power spectrum and the mean tSZ power scales approximately as $F^8$, this corresponds to a fractional uncertainty of about $8 \times 2.4 \% = 19\%$ in the reconstructed tSZ power spectrum, or a signal-to-noise ratio of 5.2, broadly consistent with the measured value of 5.6.

\begin{figure}[b]
    \centering
    \includegraphics[width=0.95\linewidth]{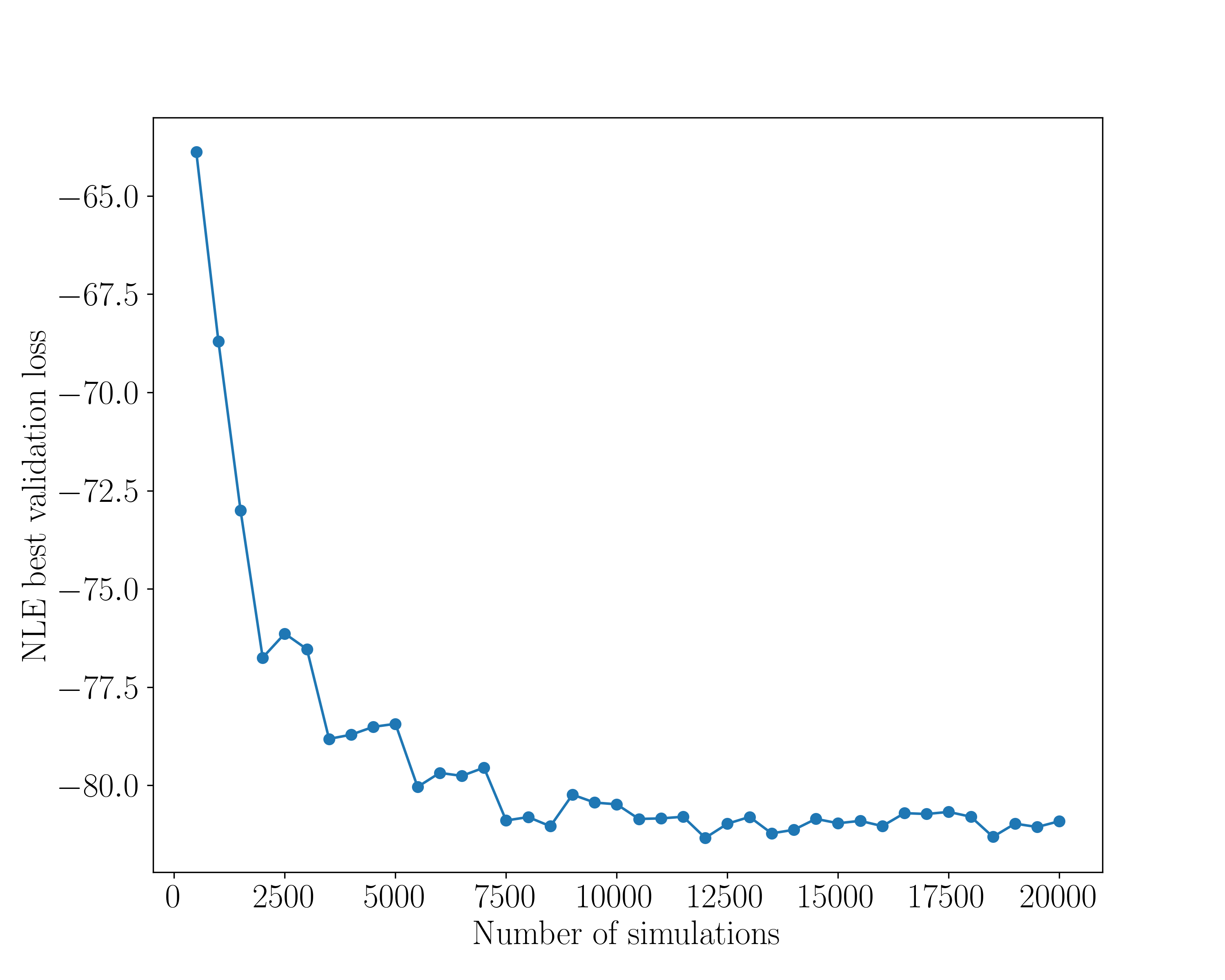}

    \vspace{0.5em}

    \includegraphics[width=0.95\linewidth]{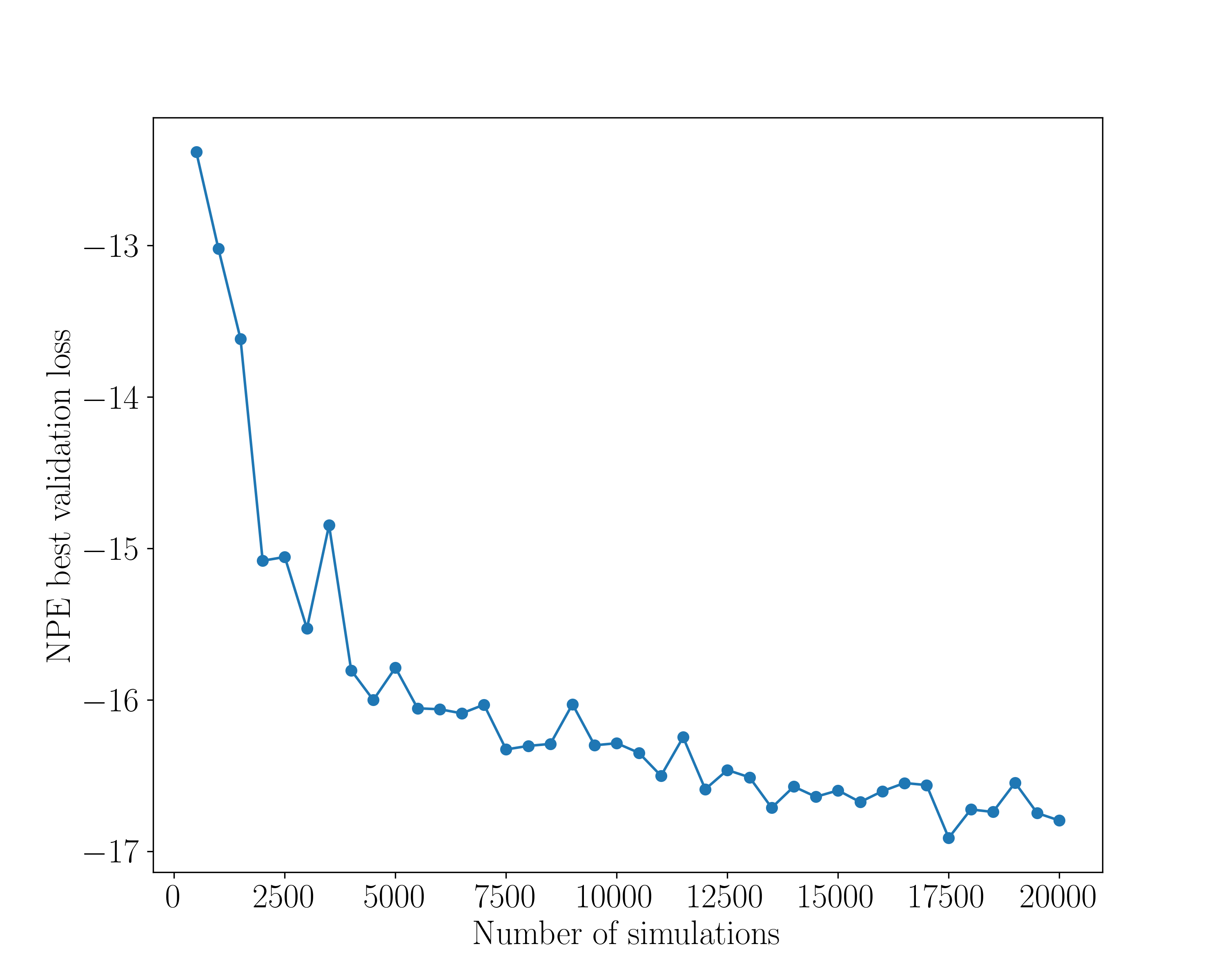}
    \caption{Best validation loss as a function of the number of training simulations. \textit{Upper panel:} NLE, converging after around $7500$ simulations. \textit{Lower panel:} NPE, converging more slowly and approaching stability after around $17{,}500$ simulations.}
    \label{fig:bestvalvstrain}
\end{figure}

\section{Validation of the SBI methods}
\label{subsec:validation_sbi}

\subsection{Convergence of SBI with number of simulations}
To assess whether the small differences in the inferred constraints for the halo-based setup (Fig.~\ref{fig:posterior_paint}) are caused by genuine non-Gaussianity in the likelihood or from training and simulation artifacts, we perform a set of ablation studies. 
We first consider the convergence of the SBI methods as a function of the number of simulations. To do so, we examine the behavior of the loss function with respect to the training sample size. The loss function is the negative log
probability used to train the neural density estimator on simulated parameter--data pairs \(\{(\boldsymbol{\theta}_i,\mathbf{t}_i)\}\), where \(\mathbf{t}\) denotes the summary statistics of the data. For NPE, this is the negative log posterior, \(-\log p_\phi(\boldsymbol{\theta}_i|\mathbf{t}_i)\), while for NLE it is the negative log likelihood, \(-\log p_\phi(\mathbf{t}_i|\boldsymbol{\theta}_i)\), where $\phi$ denotes the trainable parameters of the neural density estimator. The validation loss is
the same quantity evaluated on simulations not used for training, and therefore
provides a diagnostic of generalization.

\begin{figure*}
\includegraphics[width=0.98\linewidth]{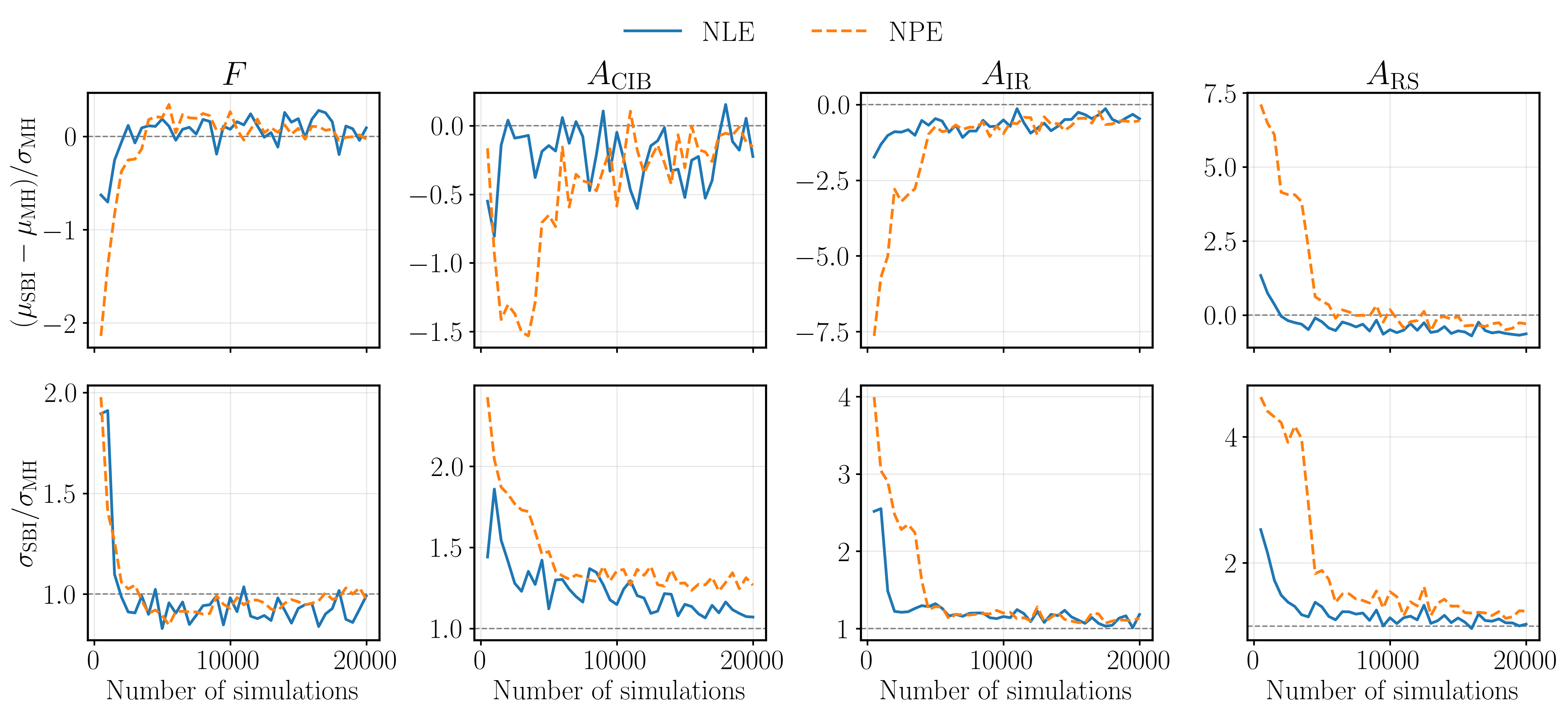}
\caption{\label{fig:comparison_bias_ratio_conv} Convergence of the posterior means (top panels) and standard deviations (bottom panels) as a function of the number of training simulations for all four parameters. Results for NLE are shown in solid blue and NPE in dashed orange. Each point represents the result from an ensemble of three independently trained neural networks. In all cases, the SBI results are referenced to the results from a Gaussian-likelihood analysis.}
\end{figure*}

Figure~\ref{fig:bestvalvstrain} shows the best validation loss as a function of the number of simulations used to train the NLE and NPE neural networks, respectively. All training runs are performed using the same hyperparameters listed in Table~\ref{tab:hyperparams} for each training set. For each training size between 500 and 20,000 simulations, we also tested different neural-network hyperparameters and found that the results are only weakly affected by this choice. For NLE, the best validation loss converges after approximately 7500 simulations, and adding additional simulations does not lead to further improvement in validation loss. NPE, however, converges more slowly; while the trend is less clear, it appears to approach convergence after around 17,500 simulations. This is possibly because more simulations are required to learn the non-Gaussian posteriors.
We further examine the convergence of the parameter means and standard deviations obtained from the posterior as the training sample size increases. Figure~\ref{fig:comparison_bias_ratio_conv} shows the convergence of the posterior means and standard deviations as a function of the number of simulations used for training. Since this process is computationally heavy (especially for NLE), we draw the posterior from an ensemble of only three independently trained neural networks for each training size (we used 10 for our earlier results). For all four parameters, the posterior means and standard deviations converge after about $N \simeq 10{,}000$ training samples. However, some residual scatter remains for individual parameters, most notably for $A_{\rm CIB}$, where fluctuations can reach up to roughly $0.5\sigma$. This may arise from several factors, such as the stochasticity in training and the finite size of the training dataset (even though the best validation loss converges). Apart from that, fluctuations in the Poisson tail of the simulated data used in training may also contribute to the observed scatter. 


\subsection{Learned likelihood (NLE)}
\label{subsec:valnle}

\begin{figure*}
\includegraphics[width=0.95\linewidth]{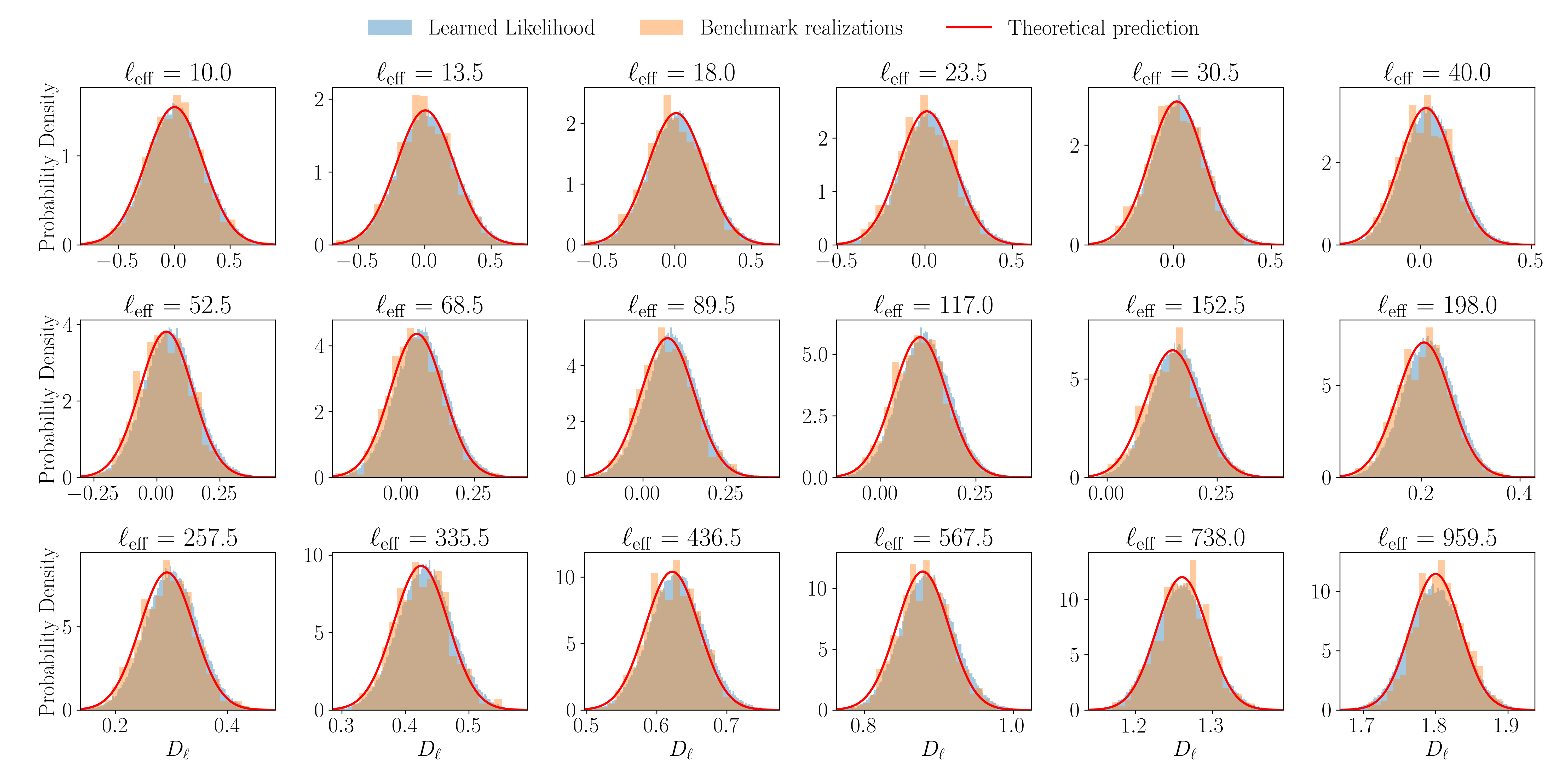}
\caption{\label{fig:lkl_hist}Histograms showing the distributions of the tSZ power spectrum drawn from the learned likelihood (blue) by assuming a Gaussian distribution at all multipoles, and the distributions of the tSZ power spectrum from 1000 different draws from the covariance matrix at the same parameter values (orange). The red curve indicates the Gaussian theoretical prediction from the mean and covariance matrix.}
\end{figure*}

One benefit of NLE is that we can validate the learned likelihood for both the Gaussian likelihood-based and halo-based SBI frameworks. We compare the learned likelihood from the SBI neural network with the actual sampling distribution of the data obtained from the simulations. 

For the SBI setup based on the Gaussian likelihood, we draw \(10^{5}\) samples
from the learned likelihood conditioned on the fiducial cosmological and residual foreground parameter values. We compare
these samples with the \(10^{3}\) independent realizations of \(D_\ell^{yy}\)
reserved as the held-out test set, generated at the same parameter values using
different noise realizations drawn from the theoretical covariance matrix. These test realizations are not used during training or validation, and provide an independent estimate of the true sampling distribution of the data vector. Figure~\ref{fig:lkl_hist} shows the resulting histograms, together with the Gaussian prediction derived from the theoretical covariance matrix. The agreement is excellent, indicating the reliability of the NLE approach in capturing the underlying distribution.


\begin{figure*}
\includegraphics[width=0.95\linewidth]{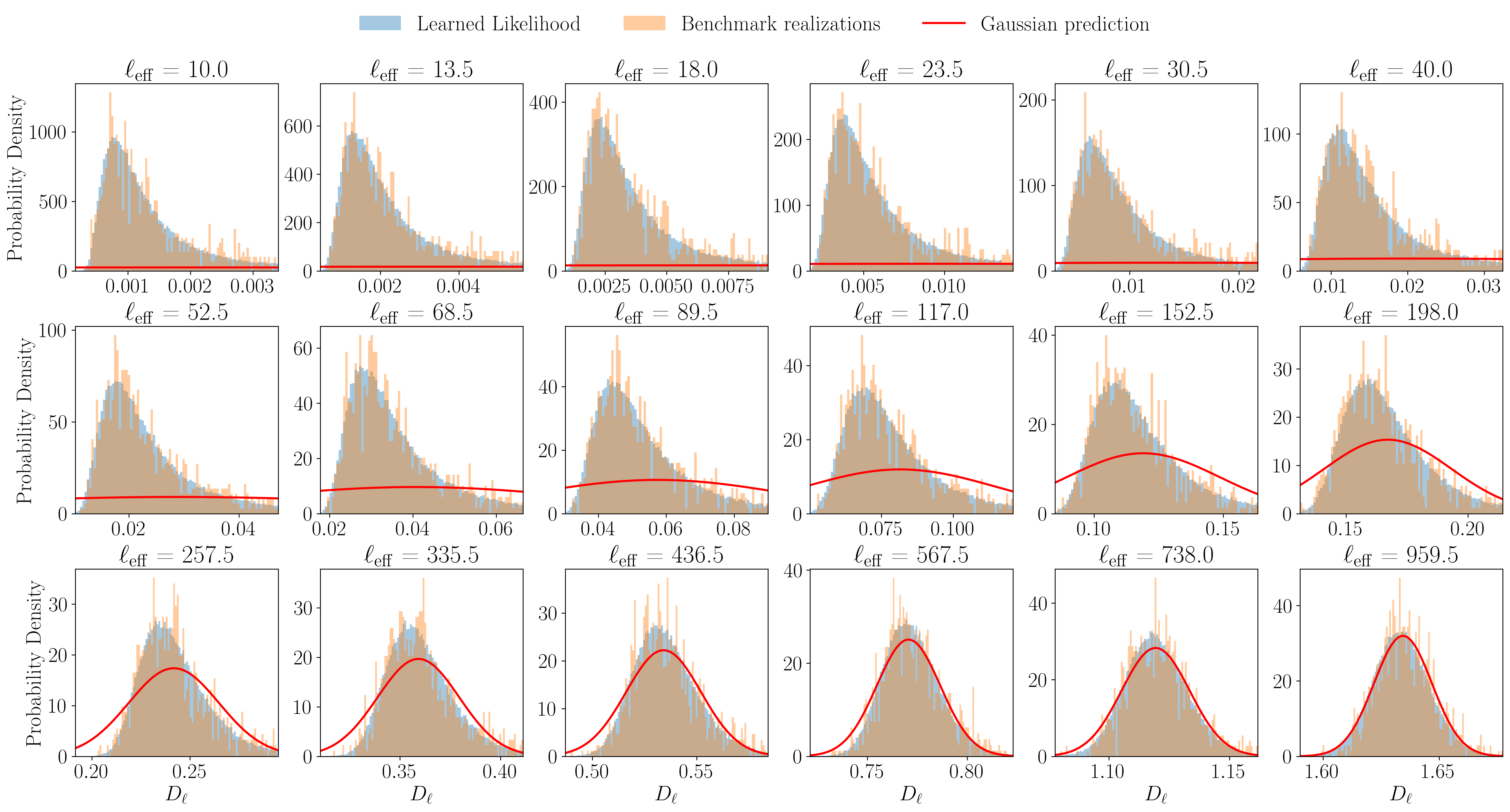}
\caption{\label{fig:paint_hist}Histograms showing the distributions of the tSZ power spectrum drawn from the learned likelihood (blue) and the sampling distributions of the tSZ power spectrum from 1000 forward-modeled realizations (orange). The red curve shows the Gaussian theoretical prediction derived from the mean and covariance matrix. At low multipoles, the Gaussian approximation fails to describe the strongly non-Gaussian distribution, whereas the learned likelihood successfully reproduces the skewness in the bandpower distribution. }
\end{figure*}

Similarly, for the halo-based approach, we perform an analogous test of drawing $10^{5}$ samples from the learned likelihood at fixed parameter values and comparing against $10^{3}$ forward-modeled realizations of $D_\ell^{yy}$. The same 1000 halo-based tSZ maps described in Sec.~\ref{subsec:halobasedsims} are used to generate these realizations. Figure~\ref{fig:paint_hist} presents the resulting histograms and the Gaussian predictions. At low multipoles, the bandpower distributions are strongly positively skewed, and the Gaussian theory curves fail to capture this highly non-Gaussian behavior. In contrast, the samples drawn from the learned NLE likelihood closely match the held-out forward-modeled distributions, including their skewness.

This comparison is also useful for diagnosing possible sources of posterior
bias. Since SBI is trained directly on the simulated data distribution, any
systematic mismatch between the forward-modeled simulations and the theoretical prediction can be inherited by the learned likelihood and hence by the inferred posterior. At higher multipoles, where foreground residuals dominate and the distribution is closer to Gaussian, we find a small mean offset of approximately \(0.1\sigma\) between the held-out forward realizations and the Gaussian prediction in bins with \(\ell>100\). This offset could therefore contribute marginally to the small shift in \(A_{\rm IR}\) seen in the SBI parameter recovery tests in Sec.~\ref{subsec:halosbi}. However, the size of the effect is too small to explain the full observed discrepancy: correcting for this offset would change the inferred \(A_{\rm IR}\) by only \(0.007\), compared to the observed shift of approximately \(0.02\).


\subsection{Coverage test (NPE)}
For the halo-based SBI case study using NPE, we perform a coverage test to assess whether the inferred posteriors are statistically consistent with the true parameters. Since NPE is amortized, the trained neural network can generate posteriors for arbitrary observations in a short time without retraining. In contrast, performing the same test for NLE is much more computationally expensive, as the posterior must be obtained by sampling from the learned likelihood for each individual observation. Moreover, the validation test for NLE described in Sec.~\ref{subsec:valnle} is already robust.
We construct 1000 independent test cases by sampling parameter sets from the prior, generating the corresponding $y$-maps and power spectra, and inferring posteriors for each using the trained network. For every test case, 10,000 posterior samples are drawn and compared to the corresponding true parameter values.
We consider two complementary diagnostics. First, we compare the posterior samples with the true input parameters across the test set, which provides a direct visual check for biases in the inferred posterior. Second, we perform a credible-level coverage test, in which we check whether posterior credible regions of nominal probability $p$ contain the true parameter values in approximately a fraction $p$ of the test simulations.

\begin{figure}[b]
\includegraphics[width=0.99\linewidth]{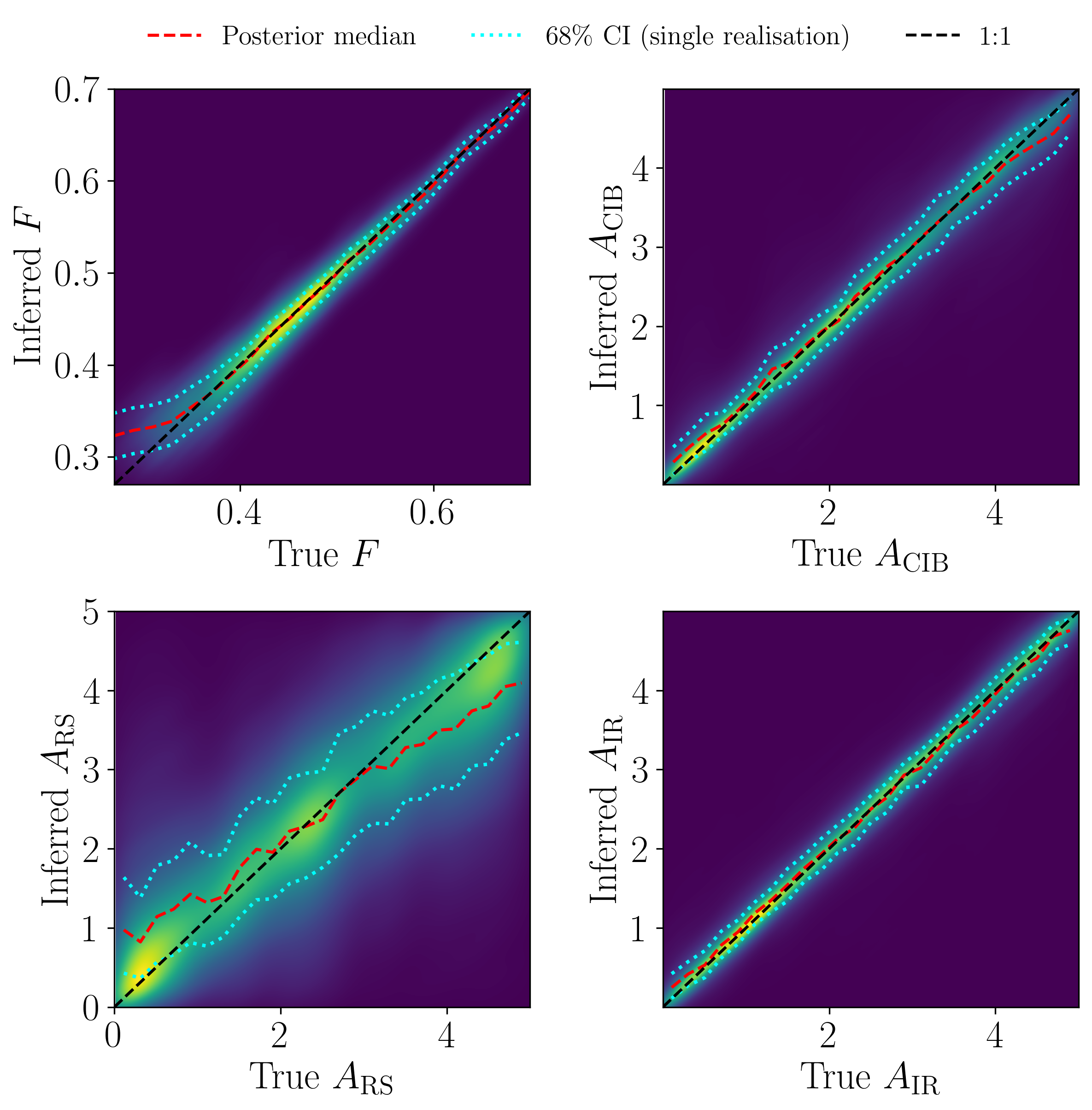}
\caption{\label{fig:coverage_plot} 
Results of the coverage test comparing the input (true) parameter values ($x$-axis) to the inferred ones ($y$-axis), using a random subsample of 100,000 points from the posterior--truth dataset. The point density is smoothed using a Gaussian kernel density estimator and shown by the color scale, with the median relation indicated by a red dashed line and the 1:1 line shown in black. The cyan dotted curves show the central 68\% credible interval for a single realization at each true parameter value, corresponding to the 16th and 84th posterior percentiles.}

\end{figure}

Figure~\ref{fig:coverage_plot} shows the two-dimensional density plots comparing the inferred posterior to the true parameter values for all four parameters. We visualize the result using a Gaussian kernel density estimator applied to a random subsample of 100,000 points from the full posterior--truth dataset. The indication that our NPE framework has good coverage is the fact that there is a high density of points around the 1:1 line for all four parameters, except at the edge of the prior. Among all four parameters, only $A_{\rm RS}$ remains poorly constrained, as expected given its weak contribution to the tSZ power spectrum. 

\begin{figure}[b]
\includegraphics[width=0.99\linewidth]{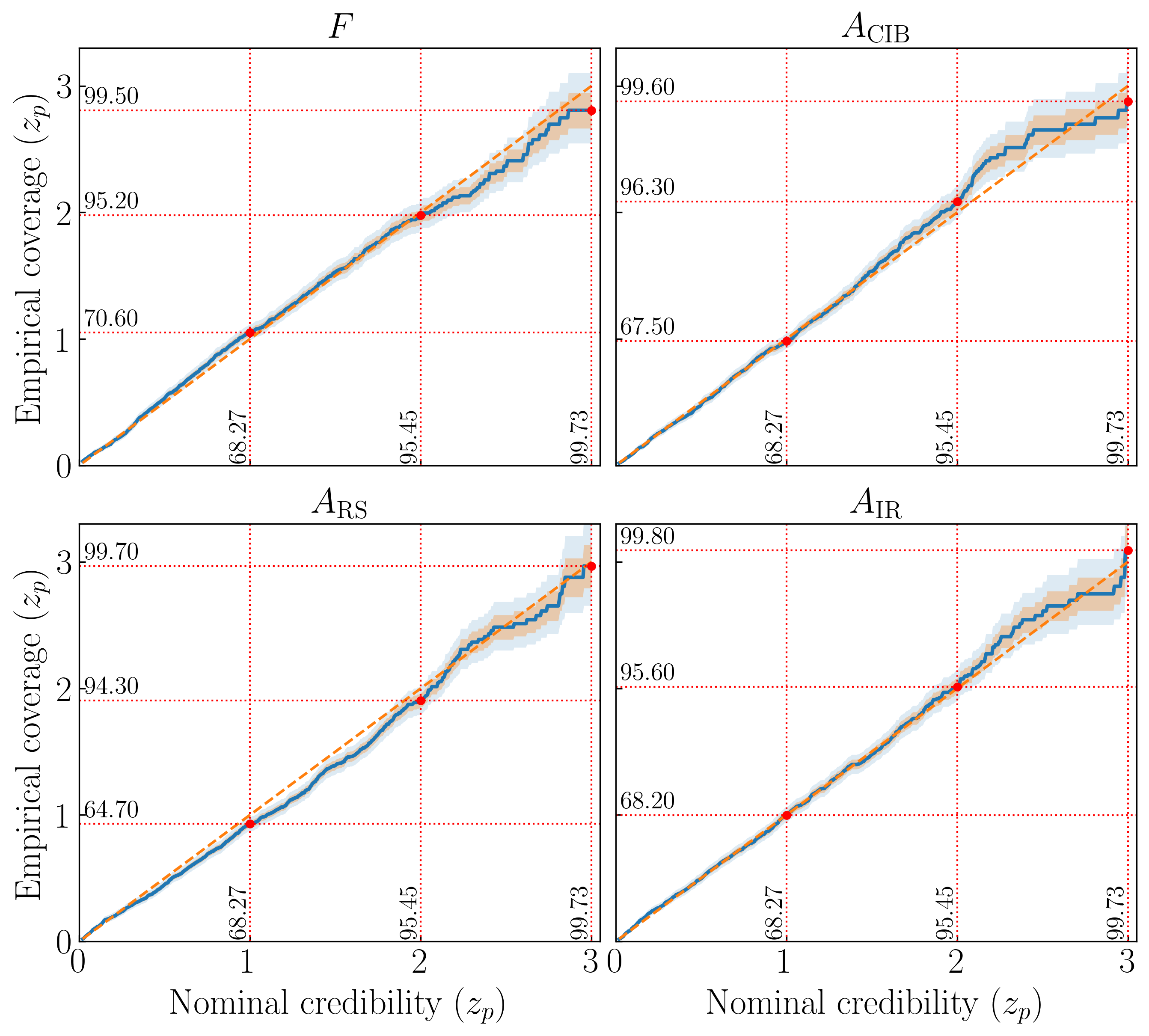}
\caption{\label{fig:credible_coverage} 
Coverage test comparing the empirical coverage to the nominal credibility level $z_p$. The dark blue lines show the mean empirical coverage, the orange shaded regions indicate the 68\% uncertainties, and the light-blue shaded regions correspond to the 95\% uncertainties. The orange dashed lines denote the 1:1 relation. The deviation near the high nominal credibility is due to the tails of the learned posterior and the finite-sample fluctuations in the test realizations.
}

\end{figure}

Fig.~\ref{fig:credible_coverage} shows the results of another coverage test. Here, for each of the posteriors corresponding to the 1000 independent prior samples and data realizations, we determine, using our trained NPE network, the highest posterior density region for each parameter enclosing probability $p/100$ (the nominal credibility\footnote{We parameterize the nominal credibility by $z_p$ through the two-sided Gaussian relation $p/100 = \frac{1}{\sqrt{2\pi}} \int_{-z_p}^{z_p}  e^{-z^2/2}\, dz$, such that $z_p = 1$, $2$, and $3$ correspond to the conventional 68.27, 95.45, and 99.73 percent credible intervals, respectively.}) and compute the fraction of these that include the true parameter value (the empirical coverage). This provides a
frequentist calibration test of the Bayesian credible regions: if the posterior
is well calibrated, the empirical coverage should agree with the nominal
credibility level, while empirical coverage below the nominal level indicates
over-confident posteriors~\citep{Hermans:2022}. The empirical coverage and the nominal credibility levels generally match to good precision. The largest deviations occur at high nominal credibility, where the test is sensitive to the tails of the learned posterior and to finite-sample fluctuations in the test realizations.


In addition, we quantify the level of bias in the posteriors by computing the ratio between the difference in the inferred mean ($\mu_{\rm SBI}$) and the true value ($\mu_{\rm true}$) and the posterior standard deviation from SBI ($\sigma_{\rm SBI}$) at those 1000 different parameter points. Table~\ref{tab:posterior_bias} summarizes this ratio for each parameter, averaged over 1000 different parameter points. For all parameters, $(\mu_{\rm SBI} - \mu_{\rm true}) / \sigma_{\rm SBI}$ is less than $0.06$, with the largest deviation being $0.0534$ for $A_{\rm IR}$.

\begin{table}[ht]
\caption{Difference between SBI-based posterior constraints and the true input parameters in units of $\sigma_{\rm SBI}$. For all parameters, the bias is less than $0.06\sigma$.}
\label{tab:posterior_bias}
\renewcommand{\arraystretch}{1.5}
\centering
\begin{ruledtabular}
\begin{tabular}{lc}
Parameter & $(\mu_{\rm SBI} - \mu_{\rm true}) / \sigma_{\rm SBI}$ \\
\colrule
$F$            & $0.0061 \pm 0.0320$ \\
$A_{\rm CIB}$  & $0.0457 \pm 0.0309$ \\
$A_{\rm RS}$   & $-0.0489 \pm 0.0327$ \\
$A_{\rm IR}$   & $0.0534 \pm 0.0311$ \\
\end{tabular}
\end{ruledtabular}
\end{table}

\section{Discussion and Outlook}
\label{sec:discussion}

In this paper, we revisited the theoretical modeling of the sampling distribution of the tSZ power spectrum, and implemented an SBI-based framework for constraining cosmological parameters from the tSZ power spectrum using halo-based simulations. We found that approximating with a Gaussian likelihood yields unbiased cosmological parameter estimates for a \textit{Planck}-like survey even in the presence of the strongly non-Gaussian sampling distribution at low multipoles. However, the constraints on the amplitudes of the foreground residuals in the SBI analysis are slightly broader than those from the Gaussian likelihood analysis, and the inferred value of the amplitude of Poisson power from infrared sources, $A_{\rm IR}$, is around $0.5\sigma$ lower relative to the likelihood-based result. We thoroughly validated the SBI implementation with a series of ablation studies in NLE and NPE to argue that this shift in $A_{\rm IR}$ is less likely due to inference systematics but rather reflects a genuine difference between the true and Gaussian likelihoods.

This is the first time that SBI has been applied to the analysis of the tSZ power spectrum. While a standard Gaussian likelihood approximation yields unbiased cosmological parameter estimates for a \textit{Planck}-like survey considered here, this conclusion does not automatically hold for other surveys such as ACT and the South Pole Telescope (SPT). ACT and SPT have lower noise and a higher angular resolution, but they both observe a smaller fraction of the sky, which increases the sample variance and makes the likelihood of the tSZ power spectrum more non-Gaussian even at higher multipoles. Moreover, the tSZ power spectrum depends more strongly on the pressure profile and baryonic effects at smaller angular scales. Even if a Gaussian likelihood is sufficient to characterize the bandpower distributions, the inference can still be limited by model misspecification in the pressure profile rather than likelihood non-Gaussianity. In this sense, SBI can offer an advantage in that it can be trained to incorporate uncertainties in the pressure profiles and small-scale astrophysical effects.

Our results also suggest that caution is required when interpreting the discrepancies between the observed tSZ power spectrum and simulation-based templates from hydrodynamical simulations, e.g.,~\cite{McCarthy:2023ism,efstathiou2025powerspectrumthermalsunyaevzeldovich}.
The bandpower distributions are positively skewed in the \textit{Planck}-like regime studied here, so the likelihood is intrinsically asymmetrical as shown in Figs~\ref{fig:tszps_reconstructed} and \ref{fig:paint_hist}. Hence, differences between measured bandpowers and a fiducial template do not map straightforwardly onto shifts in parameters. A direct visual comparison between measured bandpowers and a fiducial template can be misleading, and apparent deviations in bandpower space do not translate straightforwardly into parameter shifts.


It is worth noting that the non-Gaussianity of the tSZ power spectrum at low multipoles arises primarily from Poisson fluctuations due to a small number of local, massive clusters. Therefore, when those massive clusters are included, the tSZ power spectrum analysis is subject to large sample variance due to the trispectrum covariance. Masking these resolved clusters can therefore make the low-multipole regime more Gaussian and enhance the amount of cosmological information from the tSZ power spectrum; see, e.g.,~\cite{Shaw:2009, Osato_2021}. Meanwhile, the information in the tSZ power spectrum removed by masking is transferred to the catalog of resolved clusters. This motivates a joint analysis of the unresolved tSZ power spectrum and the number counts of the masked clusters \citep{Rotti:2020rdl}. While an analytical likelihood can provide an accurate approximation for each observable individually, the SBI framework naturally captures any covariance (especially if the large-scale clustering effect from the 2-halo term) and provides a convenient way to combine information from two different observables. We leave the further exploration of this to future work.


As a first demonstration, the present halo-based SBI implementation only considers full-sky coverage and neglects realistic noise. 
These effects can, in principle, be incorporated into the forward-modeling framework, although the accuracy of the resulting inference will depend on how well the noise properties can be realistically simulated. Extensions to incorporate realistic noise and survey masks will allow a reanalysis of current datasets such as \textit{Planck} \citep{Planck:2015vgm}, and to produce forecasts for ongoing experiments such as Simons Observatory \citep{SimonsObservatory:2018koc}. 

We finally emphasize that our forward-modeling approach is both universal and highly adaptable, providing a flexible framework for a wide range of applications beyond the specific tests presented here. While in this work we focused on the tSZ power spectrum as the summary statistic, the pipeline can equally be applied to other statistics or combined datasets. In particular, it naturally accommodates joint analyses with complementary probes such as cluster number counts \citep{Zubeldia:2025qlt}, and higher-order moments \citep{Sabyr_2025}. These extensions offer other possible directions for exploiting the full statistical power of current and forthcoming CMB surveys.

\begin{acknowledgments}
We thank William Coulton and Zack Li for assistance with the use of the \texttt{XGPaint} package. LX acknowledges the support of China Scholarship Council Cambridge Scholarship (grant number 202408060222). ÍZ and AC acknowledge support from the STFC (grant numbers ST/W000977/1, ST/X006387/1 and UKRI1164). JA is supported by a fellowship from the Kavli Foundation.

This work was performed using resources provided by the Cambridge Service for Data Driven Discovery (CSD3) operated by the University of Cambridge Research Computing Service (\texttt{www.csd3.cam.ac.uk}), provided by Dell EMC and Intel using Tier-2 funding from the Engineering and Physical Sciences Research Council (capital grant EP/T022159/1), and DiRAC funding from the Science and Technology Facilities Council (\texttt{www.dirac.ac.uk}).

Figures~\ref{fig:posterior_lkl} and \ref{fig:posterior_paint} were produced using the \texttt{getdist} package \citep{Lewis:2019xzd}.
\end{acknowledgments}

\bibliography{apssamp}
\appendix

\section{Conditional scatter of the tSZ power spectrum at fixed halo counts}
\label{app:conditional_scatter}
In Sec.~\ref{stattsz}, we model the one-point distribution of the tSZ power spectrum as arising from fluctuations in the halo counts and a residual scatter at fixed halo counts. In this appendix, we show the conditional scatter of the tSZ power spectrum at fixed halo counts is Gaussian, with variance depending on the number of halos.

Consider a fixed halo catalog, specified by the halo counts $\{N_i\}$ in sufficiently fine mass--redshift bins. Once $\{N_i\}$ is fixed, the remaining randomness in the 1-halo tSZ power spectrum comes from the angular configuration of halos on the sky. Since the tSZ field is a superposition of extended halo profiles, different angular configurations lead to different amounts of profile overlap and hence to a residual scatter in the measured power spectrum.

We expand the Compton-$y$ field in spherical harmonics,
\begin{equation}
y(\hat{\mathbf n})=\sum_{\ell m} y_{\ell m} Y_{\ell m}(\hat{\mathbf n})\, ,
\end{equation}
with
\begin{equation}
y_{\ell m}=\sum_a \tilde{y}_{\ell,a}Y_{\ell m}^*(\hat{\mathbf n}_a)\, ,
\end{equation}
where $\tilde{y}_{\ell,a}$ reduces to the two-dimensional Fourier transform of the profile of halo $a$ in the flat-sky limit, and $\hat{\mathbf n}_a$ is its sky position. The power-spectrum estimator is
\begin{equation}
\hat C_\ell=\frac{1}{2\ell+1}\sum_m |y_{\ell m}|^2\, .
\end{equation}
Using the addition theorem, this can be written as
\begin{equation}
\hat C_\ell
=
\frac{1}{4\pi}\sum_a y_{\ell,a}^2
+
\frac{1}{4\pi}\sum_{a\neq b}
y_{\ell,a}y_{\ell,b}
P_\ell(\hat{\mathbf n}_a\!\cdot\!\hat{\mathbf n}_b)\, ,
\end{equation}
where the $P_\ell$ are the Legendre polynomials. 
The first term is the self-contribution of each halo, while the second depends on the relative angular configuration of distinct halos. The latter therefore describes the additional scatter induced by the overlap of halo profiles at fixed halo catalog.

Averaging only over the halo positions while keeping the catalog fixed, statistical isotropy implies
\begin{equation}
\Big\langle
P_\ell(\hat{\mathbf n}_a\!\cdot\!\hat{\mathbf n}_b)
\Big\rangle_{\{N_i\}}
=0\, ,
\qquad a\neq b\, ,
\end{equation}
for $\ell>0$, and hence
\begin{equation}
\langle \hat C_\ell\rangle_{\{N_i\}}
=
\frac{1}{4\pi}\sum_a y_{\ell,a}^2
\equiv
\bar C_\ell(\{N_i\})\, .
\end{equation}
In binning notation, this becomes
\begin{equation}
\bar C_\ell(\{N_i\})
=
\frac{1}{\Omega_{\rm sky}}
\sum_i N_i S_i\, ,
\end{equation}
where $S_i=|\tilde y_\ell(M_i,z_i)|^2$.

The fluctuation about this conditional mean is therefore
\begin{equation}
\delta\hat C_\ell
\equiv
\hat C_\ell-\bar C_\ell(\{N_i\})
=
\frac{1}{4\pi}\sum_{a\neq b}
y_{\ell,a}y_{\ell,b}
P_\ell(\hat{\mathbf n}_a\!\cdot\!\hat{\mathbf n}_b)\, ,
\end{equation}
so the residual scatter at fixed halo counts is sourced by the off-diagonal pair terms.
The conditional covariance of the $\delta \hat{C}_\ell$ is
\begin{align}
\text{cov}(\delta\hat{C}_\ell , \delta \hat{C}_{\ell'})_{\{N_i\}}
&= \frac{1}{(4\pi)^2}\sum_{a\neq b}\sum_{c\neq d} \tilde{y}_{\ell,a} \tilde{y}_{\ell,b} \tilde{y}_{\ell',c} \tilde{y}_{\ell',d} \nonumber\\
&\quad\times \langle P_\ell(\hat{\mathbf{n}}_a \cdot \hat{\mathbf{n}}_b) P_{\ell'}(\hat{\mathbf{n}}_c \cdot \hat{\mathbf{n}}_d)\rangle_{{\{N_i\}}} \,.
\end{align}
The average over halo locations vanishes for $\ell\neq 0$ and $\ell'\neq 0$ unless the sum over four halos reduces to two pairs, i.e., $a=c$ and $b=d$ or $a=d$ and $b=c$, which contribute equally. Using the addition theorem, we have
\begin{equation}
\langle P_\ell(\hat{\mathbf{n}}_a \cdot \hat{\mathbf{n}}_b) P_{\ell'}(\hat{\mathbf{n}}_a \cdot \hat{\mathbf{n}}_b)\rangle_{{\{N_i\}}} = \frac{1}{2\ell+1} \delta_{\ell\ell'} \,, \qquad a\neq b\, ,
\end{equation}
so that
\begin{equation}
\text{cov}(\delta\hat{C}_\ell , \delta \hat{C}_{\ell'})_{\{N_i\}} = \frac{2 \delta_{\ell \ell'}}{2\ell+1} \frac{1}{(4\pi)^2} \sum_{a\neq b} \tilde{y}^2_{\ell,a} \tilde{y}^2_{\ell,b} \, . \label{eq:covdeltaCl}
\end{equation}

For a large number of clusters, the sum of the diagonal $a=b$ terms that are excluded from the covariance in Eq.~\eqref{eq:covdeltaCl} is small in comparison to those terms retained. Adding back in the diagonal terms, we can therefore approximate
\begin{equation}
 \text{cov}(\delta\hat{C}_\ell , \delta \hat{C}_{\ell'})_{\{N_i\}}  \approx \frac{2 \delta_{\ell \ell'}}{2\ell+1}\bar{C}^2_\ell(\{N_i\}) \, .
\end{equation}
This is simply the disconnected (Gaussian) variance for the estimated power spectrum of a field with true power spectrum $\bar{C}_\ell(\{N_i\})$. If we further approximate the conditional fluctuations $\delta\hat{C}_\ell$ as being Gaussian distributed, the conditional bandpower distribution may be approximated as
\begin{equation}
\begin{split}
P(C\mid\{N_i\})
&\approx \frac{1}{\sqrt{2\pi\sigma_G^2(\{N_i\})}} \\
&\quad\times
\exp\!\left[
-\frac{\bigl(C-\bar C(\{N_i\})\bigr)^2}{2\sigma_G^2(\{N_i\})}
\right] \,.
\end{split}
\end{equation}
where, after multipole binning and accounting for partial sky coverage, the bandpower variance is
\begin{equation}
\sigma_G^2(\{N_i\})
\approx
\frac{2\,\bar C_\ell^2(\{N_i\})}{(2\ell+1)\Delta\ell\,f_{\rm sky}} \, .
\label{eq:appbandpowerdist}
\end{equation}

Equation~\eqref{eq:appbandpowerdist} shows that the conditional Gaussian variance is, strictly speaking, a function of the halo catalog. In the main text, we neglect this mild dependence and evaluate it at the mean catalog, or equivalently at a typical power-spectrum amplitude. In particular, if we return to the exact expression~\eqref{eq:covdeltaCl} and average over the halo catalog, we have
\begin{equation}
\text{cov}(\delta\hat{C}_\ell , \delta \hat{C}_{\ell'}) = 
\frac{2 \delta_{\ell \ell'}}{2\ell+1}\left(C^{\text{tSZ,\,1h}}_\ell\right)^2 \, .
\label{eq:fullcovdeltaCl}
\end{equation}
This follows from the rearrangement
\begin{align}
\Bigl\langle \sum_{a\neq b} \tilde{y}^2_{\ell,a} \tilde{y}^2_{\ell,b}\Bigr\rangle &= \Bigl\langle \sum_{ab} \tilde{y}^2_{\ell,a} \tilde{y}^2_{\ell,b}\Bigr\rangle - \Bigl\langle \sum_{a} \tilde{y}^4_{\ell,a} \Bigr\rangle \nonumber \\
&\rightarrow \Bigl\langle \sum_{i j} N_i N_j S_i S_j \Bigr\rangle - \Bigl\langle \sum_{i} N_i S_i^2 \Bigr\rangle \nonumber \\
&= \sum_{ij} \lambda_i \lambda_j S_i S_j \nonumber \\
&= \left(C^{\text{tSZ,\,1h}}_\ell\right)^2 \, ,
\end{align}
where $\lambda_i = \langle N_i\rangle$ is the mean number of halos in mass and redshift bin $i$, and we have assumed Poisson statistics for the $\{N_i\}$. Equation~\eqref{eq:fullcovdeltaCl} is simply the disconnected (Gaussian) covariance for a field with power spectrum $C^{\text{tSZ,\,1h}}_\ell$. The full covariance of $\hat{C}_\ell$ has an additional contribution from $\text{cov}\left(\bar{C}_\ell(\{N_i\}), \bar{C}_{\ell'}(\{N_i\})\right)$, which is exactly the trispectrum term:
\begin{equation}
\begin{split}
 \text{cov}\left(\bar{C}_\ell(\{N_i\}), \bar{C}_{\ell'}(\{N_i\})\right)
 &= \frac{1}{(4\pi)^2} \sum_i \lambda_i |\tilde{y}_\ell(M_i,z_i)|^2 \\
 &\quad \times |\tilde{y}_{\ell'}(M_i,z_i)|^2 \,.
\end{split}
\end{equation}

In the Planck-like regime considered in this paper, the dominant non-Gaussianity of the tSZ power spectrum still arises from fluctuations in the abundance of rare massive halos, while the additional scatter at fixed halo counts is subdominant. A more complete derivation in the halo-model framework is given by Ref.~\cite{Atkins2025CertainUncertainties}.

\section{Details of sky map generation}
\subsection{Generating cluster catalogs}
\label{app:gencatprecision}
\begin{figure*}[!t]
\includegraphics[width=0.99\linewidth]{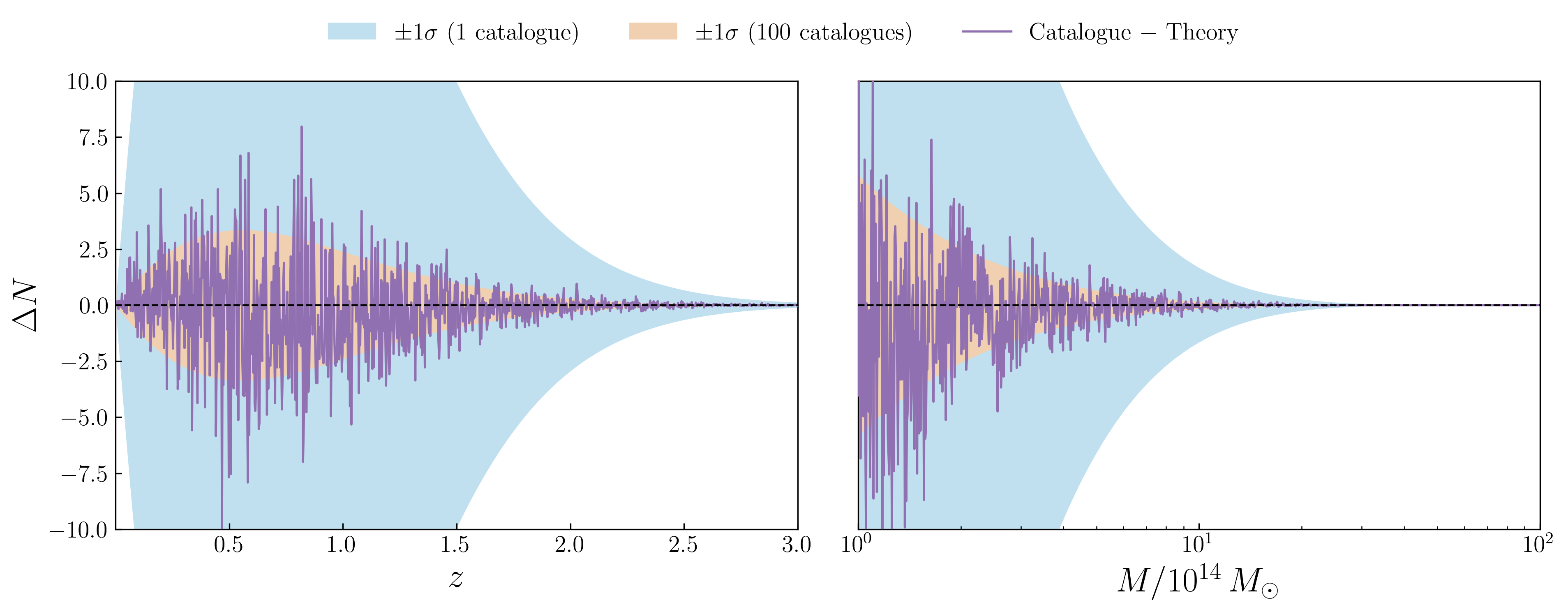}
\caption{\label{fig:cat_hist} 
Residuals between the mean number counts and the theoretical prediction (purple), shown with Poisson error bars for a single catalog (blue) and for 100 catalogs (orange). From the coverage statistics, the synthetic catalogs agree well with the theoretical predictions.}
\end{figure*}
We assess the agreement between our synthetic cluster catalogs and the corresponding theoretical predictions. The catalogs are constructed with a minimum halo mass of $M_{\rm min} = 10^{14} M_\odot$ and a maximum mass of $M_{\rm max} = 10^{16} M_\odot$, spanning a redshift range of $z_{\rm min} = 0.005$ to $z_{\rm max} = 3.0$. Figure~\ref{fig:cat_hist} shows the difference between the theoretical prediction and the mean cluster number counts as functions of redshift and mass, computed by averaging over 100 synthetic catalogs using precision parameters $n_M = 5000$ and $n_z = 5000$, where $n_M$ and $n_z$ are the number of mass and redshift bins we used to compute HMF. Since the tSZ power spectrum is particularly sensitive to the abundance of low-redshift local clusters, this choice of precision parameters ensures that the catalog generation correctly reproduces the expected cluster abundance in the regime most relevant for the power-spectrum analysis. The synthetic catalogs show good agreement with the theoretical predictions. For the number counts binned in redshift, 69.1\% lie within the expected $\pm1\sigma$ Poisson uncertainty, while for the counts binned in mass, 76.8\% fall within the same range. These coverage statistics indicate that the synthetic catalogs are consistent with the theoretical predictions in cluster number counts. 



\subsection{Painting clusters onto sky maps}
Once the catalogs are generated, each cluster is assigned a beam-convolved Compton-$y$ profile based on its mass, redshift, and assumed pressure profile. The clusters are then projected onto a full-sky \texttt{HEALPix} grid to produce the synthetic $y$-maps. We additionally implement the GNFW pressure profile, parameterized by \citet{Arnaud_2010}, on top of the default pressure profile, given by \citet{Battaglia_2012}, in \texttt{XGPaint}. In general, the pressure profile, $y(M,z,\theta)$, depends on two parameters, $M$ and $z$, and the angular coordinate $\theta$. One might build a three-dimensional interpolation grid to compute the pressure profile at any given mass, redshift and angle. This would need to be reconstructed every time the cosmological parameters are varied. However, the GNFW pressure profile with a constant concentration parameter, $c_{500}$, reduces this to an effectively one-dimensional problem. Specifically, the GNFW pressure profile can be rewritten as $y(M,z,\theta) = y_0(M,z)y_t(\theta/\theta_{500})$, where $y_0$ is the amplitude of the Compton-$y$ parameter evaluated at the cluster center, and $y_t$ is the spatial template of the cluster. In our case, $y_0$ and $y_t$ can be analytically expressed as
\begin{equation}
\label{eq:y0}
y_0 = \frac{2\sigma_{\text{T}}}{m_e c^2} R_{500} P_{500}
\, I\!\left(\frac{\theta}{\theta_{500}} = 0\right) \, ,
\end{equation}
where
\begin{widetext}
\begin{equation}
\label{eq:P500}
P_{500} = 1.65 \left(\frac{H(z)}{H_0}\right)^{8/3} P_0 \left[\frac{M_{500}}{3\times 10^{14}\, h_{70}^{-1}\, M_\odot}\right]^{2/3 + 0.12} \left(\frac{h}{0.7}\right)^{2} \;\mathrm{eV\,cm^{-3}} \, ,
\end{equation}
\noindent and
\begin{equation}
\label{eq:I_theta}
I\!\left(\frac{\theta}{\theta_{500}}\right) = \int_0^\infty \mathrm{d}u\, \Big[ c_{500}\, \sqrt{u^2 + \left( \frac{\theta}{\theta_{500}} \right)^2}\,\Big]^{-\gamma} \Bigg\{  1 + \Big[ c_{500}\, \sqrt{u^2 + \left( \frac{\theta}{\theta_{500}} \right)^2}\,\Big]^{\alpha} \Bigg\}^{\frac{\gamma - \beta}{\alpha}} \, .
\end{equation}
\end{widetext}
and $y_t = I(\theta/\theta_{500})/I(0)$. To include beam convolution, we construct a two-dimensional interpolation grid in $\theta$ and $\theta_{500}$ that is fast and more accurate with high-precision parameters of $N_{\rm log\theta}=N_{\rm log\theta_{500}}=8192$. To perform the beam convolution, we use the Fast Hankel Transform (FHT) method in \texttt{Pixell.jl} to convolve the profiles with the beam with the same precision parameter on the $\theta$-grid. We choose the real-space angular range to be $\log\theta_{\min}=-16.5$ and $\log\theta_{\max}=2.5$, which corresponds to an $\ell$ range of [0.082, $1.5 \times10^7$].

\end{document}